\pdfoutput=1
\documentclass[aps,prd,preprint,nofootinbib]{revtex4-1}
\usepackage{graphicx}
\usepackage{hyperref}
\usepackage{amsmath}
\usepackage{color}

\usepackage{etoolbox}
\apptocmd{\thebibliography}{\interlinepenalty=10000}{}{}

\AtBeginDocument{\count\footins=1000}

\providecommand{\color}[1]{}

\begin{document}

\title{Zee Model with Flavor Dependent Global $U(1)$ Symmetry}

\author{Takaaki Nomura}
\email{nomura@kias.re.kr}
\affiliation{School of Physics, KIAS, Seoul, 02455, Korea}

\author{Kei Yagyu}
\email{yagyu@het.phys.sci.osaka-u.ac.jp}
\affiliation{Department of Physics, Osaka University, Toyonaka, Osaka 560-0043, Japan}

\preprint{KIAS-P19030}
\preprint{OU-HET 1013}

\begin{abstract}
 
We study a simple extension of the Zee model, in which a discrete $Z_2$ symmetry imposed in the original model 
is replaced by a global $U(1)$ symmetry retaining the same particle content. 
Due to the $U(1)$ symmetry with flavor dependent charge assignments, 
the lepton sector has an additional source of flavor violating Yukawa interactions with a controllable structure, while the quark sector does not at tree level. 
We show that current neutrino oscillation data can be explained under constraints from lepton flavor violating decays of charged leptons in a successful charge assignment of the $U(1)$ symmetry.  
In such scenario, we find a characteristic pattern of lepton flavor violating decays of additional Higgs bosons, which can be a smoking gun signature at collider experiments.  

\end{abstract}

\maketitle

\section{Introduction}

The observation of neutrino oscillations has been known as one of clear evidence of the existence of new physics beyond the Standard Model (SM). 
It suggests that neutrinos have nonzero masses of order 0.1 eV which is remarkably smaller than the other known fermion masses. 
This fact leads us to think that neutrino masses are generated by a mechanism different from those for SM charged fermions, e.g., 
Majorana masses introduced by a lepton number violation. 

The type-I seesaw mechanism~\cite{Minkowski:1977sc,Yanagida:1980xy,Mohapatra:1979ia} provides an excellently simple explanation for tiny neutrino masses and their mixings 
just by introducing right-handed neutrinos into the SM. 
Taking the mass of right-handed neutrinos to be ${\cal O}(10^{14{\rm -}15})$ GeV with ${\cal O}(1)$ Dirac Yukawa couplings, one can 
reproduce the correct order of the neutrino mass. 
Despite such simpleness, it has been argued that a heavy right-handed neutrino is quite challenging to detect at collider experiments. 

As an alternative scenario, 
so-called radiative neutrino mass models can naturally explain tiny neutrino masses without introducing super heavy particles thanks to loop suppression factors.  
The model by A.~Zee~\cite{Zee:1980ai} proposed in 80's is the first one, in which only the scalar sector is extended from the SM, containing two isospin doublets and a charged singlet scalar fields. 
The lepton number violation is introduced via scalar interactions, and then neutrino masses are generated at one-loop level. 
After the Zee model appeared, various versions of the radiative neutrino mass models were proposed, 
and some of them can also explain the existence of dark matter~\cite{Krauss:2002px,Ma:2006km} 
and the baryon asymmetry of the Universe~\cite{Aoki:2008av}. 

Although the Zee model gives a simple and natural explanation for the smallness of neutrino masses, the original model cannot accommodate 
neutrino mixing data, because it has the too constrained structure of lepton flavor violating (LFV) Yukawa couplings~\cite{Koide:2001xy,Frampton:2001eu,He:2003ih}. 
Namely, the LFV interactions only come from an anti-symmetric $3\times 3$ matrix for the coupling among the charged singlet scalar and lepton doublets. 
One can add another source of flavor violating interactions if both two doublet Higgs fields are allowed to couple with charged leptons, known as 
the Type-III Yukawa interaction of two Higgs doublet models (THDMs)~\cite{He:2003ih,AristizabalSierra:2006ri,He:2011hs,Herrero-Garcia:2017xdu}. 
As the other directions of the extension, 
$A_4$ symmetric~\cite{Fukuyama:2010ff} and supersymmetric~\cite{Kanemura:2015maa} versions of the Zee model have also been discussed.

The extension of the Zee model with the Type-III Yukawa interaction can actually explain current neutrino data, see e.g.,~\cite{Herrero-Garcia:2017xdu}. 
However, the quark sector should also be expected to have flavor violating Yukawa interactions as these are generally allowed by the symmetry. 
Thus, in order to avoid dangerous flavor changing neutral currents (FCNCs) for the quark sector, one needs to tune quark Yukawa couplings by hand, i.e., most of their off-diagonal elements 
have to be taken to extremely small. 
In addition, there appear too many parameters in the lepton Yukawa couplings, i.e., totally 42 degrees of freedom in general (cf. 24 degrees of freedom in the original model), which 
makes the theory less predictive. 

In this paper, we would like to simultaneously overcome the above mentioned shortcomings in the Zee model with the Type-III Yukawa interaction. 
Our approach is quite simple.
Namely, we just replace a discrete $Z_2$ symmetry imposed in the original model with a global $U(1)$ symmetry with a flavor dependent charge assignment in the lepton sector.  
By taking an appropriate assignment, 
we can obtain an additional source of the lepton flavor violation whose structure is controllable by the symmetry.
On the other hand, the structure of the quark sector remains the same as that in the original model. 
We then can successfully solve the two problems in the model with the Type-III Yukawa interaction. 
In fact, models with this kind of a flavor dependent global $U(1)$ symmetry have been known as the Branco-Grimus-Lavoura (BGL) model~\cite{Branco:1996bq} (for the recent work, see e.g.,~\cite{Alves:2018kjr}), 
in which the global $U(1)$ symmetry has been originally imposed to the quark sector. 
Thus, our approach corresponds to the application of the BGL model to the lepton sector. 
We find that there are parameter sets to explain current neutrino data under the constraint from LFV processes in the model with 
an appropriate charge assignments of the $U(1)$ symmetry. 
We then clarify that a characteristic pattern of LFV decays of additional Higgs bosons is predicted, which can be a smoking gun signature to test our model at collider experiments. 

This paper is organized as follows. 
In Sec.~\ref{sec:model}, we define our model. We then give the expressions for the Yukawa interactions and the Higgs potential. 
In Sec.~\ref{sec:numass}, we calculate neutrino mass matrix which is generated at one-loop level, and discuss the constraint from LFV decays of charged leptons. 
Sec.~\ref{sec:pheno} is devoted for numerical evaluations of LFV decays of charged leptons and additional Higgs bosons. 
Conclusions are given in Sec.~\ref{sec:conc}. 
In Appendix~\ref{sec:class}, we define three classes for the $U(1)$ charge assignments, and show the structure of lepton Yukawa matrices for each class. 
In Appendix~\ref{sec:mass}, the formulae for the scalar boson masses and mixings are presented. 
In Appendix~\ref{sec:LFV_amp}, we give the analytic expressions for the amplitudes of LFV decays of charged leptons. 

\section{Model}\label{sec:model}

\begin{table}[!h]
\begin{center}
\begin{tabular}{|c||cccccccc|}\hline
          & $Q_L^i$ & $u_R^i$ & $d_R^i$ & $L_L^i$ & $\ell_R^i$ & $\Phi_1$ & $\Phi_2$  & $S^+$ \\\hline\hline
$SU(3)_c$ & ${\bf 3}$  & ${\bf 3}$ & ${\bf 3}$ & ${\bf 1}$  & ${\bf 1}$ & ${\bf 1}$ & ${\bf 1}$ & ${\bf 1}$  \\\hline
$SU(2)_L$ & ${\bf 2}$  & ${\bf 1}$ & ${\bf 1}$ & ${\bf 2}$  & ${\bf 1}$ & ${\bf 2}$ & ${\bf 2}$ & ${\bf 1}$   \\\hline
$U(1)_Y$             & $1/6$      & $2/3$     &  $-1/3$   &  $-1/2$    &  $-1$  &  $1/2$ &  $1/2$ &  $1$                 \\\hline
$U(1)'$   & 0          & 0         & 0    &   $q_L^i$  &  $q_R^i$   &  $q$   &  0     &  $q_S$ \\\hline
\end{tabular}
\caption{Particle content and charge assignment under the SM gauge symmetry $SU(3)_c \times SU(2)_L \times U(1)_Y$ and the additional 
global $U(1)'$ symmetry. The $U(1)'$ charges $q_L^i$ and $q_R^i$ are flavor dependent. }
\label{particle}
\end{center}
\end{table}

The particle content is the same as that of the original Zee model~\cite{Zee:1980ai} shown as in Table~\ref{particle}. 
In this table, $Q_L^i$ ($L_L^i$) are the left-handed quark (lepton) doublets, 
while $u_R^i$, $d_R^i$ and $\ell_R^i$ are respectively the right-handed up-type, down-type quarks and charged lepton singlets.
The superscript $i=1,2,3$ denotes the flavor index. 
The scalar sector is extended from the minimal form assumed in the SM, which is composed of two isospin doublet Higgs fields $\Phi_{1,2}$ and a charged singlet scalar field $S^\pm$. 

In the original model, a softly-broken discrete $Z_2$ symmetry is introduced, by which only one of two Higgs doublets couples to each type of fermions. 
Thus, the quark sector does not have FCNCs mediated by neutral Higgs bosons at tree level. 
On the other hand, the source of lepton flavor violation is only induced via Yukawa interactions with $S^\pm$, and it is not enough to explain the current neutrino mixing data~\cite{Koide:2001xy,Frampton:2001eu,He:2003ih}. 
In our model, we introduce a global $U(1)$ symmetry, denoting $U(1)'$, instead of the $Z_2$ symmetry. 
Charge assignments for the $U(1)'$ symmetry are given in Table~\ref{particle}. 
Because the charges for the left-handed leptons $q_L^i$ and the right-handed leptons $q_R^i$ are flavor dependent, 
we obtain another source of the lepton flavor violation 
from Yukawa interactions for the Higgs doublets. 
On the contrary, quark fields are not charged under $U(1)'$, so that the quark Yukawa interaction remains as the original form. 
In order to avoid an undesired massless Nambu-Goldstone (NG) boson associated with the spontaneous breaking of $U(1)'$, 
we introduce explicit and soft breaking terms of the $U(1)'$ symmetry in the Higgs potential. 
We note that our $U(1)'$ symmetry is anomalous in the sense that the left- and right-handed leptons are charged differently, which does not 
cause any theoretical and phenomenological problems. 
We also note that the $U(1)'$ symmetry can be replaced by a discrete symmetry by taking appropriate charge assignments, see Appendix~\ref{sec:class}. 

In this section, we first construct the Lagrangian for the Yukawa interaction in Sec.~\ref{sec:yukawa}, and then 
discuss the Higgs potential in Sec.~\ref{sec:potential}. 

\subsection{Yukawa interaction\label{sec:yukawa}}

The most general form of the Yukawa Lagrangian is given by 
\begin{align}
\mathcal{L}_Y & =
-(\tilde{Y}_u)_{ij} \bar{Q}_L^i \Phi_2^c u_R^j -(\tilde{Y}_d)_{ij} \bar{Q}_L^i \Phi_2 d_R^j + \text{h.c.}\notag\\
&-(\tilde{Y}_\ell^1)_{ij} \bar{L}_L^i \Phi_1 \ell_R^j -(\tilde{Y}_\ell^2)_{ij} \bar{L}_L^i \Phi_2 \ell_R^j
-\tilde{F}_{ij}\overline{L_L^{ci}}(i\tau_2)L_L^jS^+ + \text{h.c.}, 
\end{align}
where $\tilde{Y}_u$ and $\tilde{Y}_d$ are general complex $3\times 3$ Yukawa matrices, while the structure of $\tilde{Y}_{\ell}^{1,2}$ and $\tilde{F}$ 
depends on the $U(1)'$ charges. We note that regardless of the charge assignments, $\tilde{F}$ is the anti-symmetric $3\times 3$ matrix, because 
the $SU(2)_L$ index for the lepton doublets is contracted by the anti-symmetric way. 
The concrete structure of $\tilde{Y}_\ell^{1,2}$ and $\tilde{F}$ is presented in Appendix~\ref{sec:class}. 
In the above expression, fields with the superscript $c$ denote their charge conjugated one. 

In order to separately write the fermion mass term and the other interaction terms, we introduce the Higgs basis defined as 
\begin{align}
\begin{pmatrix}
\Phi_1 \\
\Phi_2 
\end{pmatrix}
=\begin{pmatrix}
c_\beta & -s_\beta \\
s_\beta & c_\beta
\end{pmatrix}
\begin{pmatrix}
\Phi \\
\Phi' 
\end{pmatrix}, ~~\text{with}~~
\Phi = 
\begin{pmatrix}
G^+ \\
\frac{h_1' + v + iG^0 }{\sqrt{2}}
\end{pmatrix},\quad 
\Phi' = 
\begin{pmatrix}
H^+ \\
\frac{h_2'  + iA }{\sqrt{2}}
\end{pmatrix}, \label{eq:phi}
\end{align}
where $s_X^{} = \sin X$ and $c_X^{} = \cos X$. 
The mixing angle $\beta$ is determined by the ratio of the vacuum expectation values (VEVs), i.e.,  $\tan\beta = \langle \Phi_2^0\rangle /\langle \Phi_1^0\rangle$ 
with $\Phi_{1,2}^0$ being the neutral component of the Higgs doublets and $v^2 = 2(\langle \Phi_1^0\rangle^2 + \langle \Phi_2^0\rangle^2) \simeq (246$ GeV$)^2$. 
In Eq.~(\ref{eq:phi}), $G^\pm$ and $G^0$ are NG bosons which are absorbed into the longitudinal components of $W$ and $Z$ bosons, respectively, while 
$H^\pm$, $h_{1,2}'$ and $A$ are physical charged, CP-even and CP-odd Higgs bosons, respectively. 
We note that our model does not contain physical CP-violating phases in the Higgs potential, because the $(\Phi_1^\dagger \Phi_2)^2$ term is forbidden by the $U(1)'$ symmetry, 
as it will be clarified by looking at the explicit form of the Higgs potential given below. 
potential 
Among these physical Higgs bosons, the CP-even Higgs bosons and a pair of charged Higgs bosons $H^\pm$ are not mass eigenstates in general, where the latter can mix with the singlet scalar $S^\pm$. 
Their mass eigenstates are defined as 
\begin{align}
\begin{pmatrix}
h_1' \\
h_2'
\end{pmatrix}
&=\begin{pmatrix}
c_{\alpha -\beta} & -s_{\alpha -\beta} \\
s_{\alpha -\beta} & c_{\alpha -\beta}
\end{pmatrix}
\begin{pmatrix}
H\\
h
\end{pmatrix}, \quad 
\begin{pmatrix}
H^\pm \\
S^\pm
\end{pmatrix}
=\begin{pmatrix}
c_\chi & -s_\chi \\
s_\chi & c_\chi
\end{pmatrix}
\begin{pmatrix}
H_1^\pm\\
H_2^\pm
\end{pmatrix}, 
\end{align}
where the mixing angles $\alpha$ and $\chi$ are expressed in terms of the parameters in the Higgs potential.  
We identify $h$ as the discovered Higgs boson with a mass of about 125 GeV.  

The Yukawa Lagrangian is then rewritten in the Higgs basis as 
\begin{align}
\mathcal{L}_Y & =
- \frac{\sqrt{2}m_u}{v}
(\bar{u}_L',\bar{d}_L'V^\dagger )
  (\Phi^c + \cot \beta \Phi^{c\prime }) u_R^{\prime} 
-  \frac{\sqrt{2}m_d}{v} (\bar{u}_L'V ,\bar{d}_L' ) (\Phi + \cot \beta \Phi' ) d_R^{\prime} + \text{h.c.}\notag\\
&-(\bar{\nu}_L,\bar{\ell}_L') \left[ 
\frac{\sqrt{2}}{v}\begin{pmatrix}
M_\ell U_R\,  G^+ \\
m_\ell   \,   \Phi^0
\end{pmatrix} 
 + 
\begin{pmatrix}
Y_\ell  \,H^+ \\
Y_\ell^0  \, \Phi^{\prime 0}
\end{pmatrix} 
  \right] \ell_R^{\prime} 
 -\tilde{F} \overline{L_L^c}     (i\tau_2)L_L S^+ + \text{h.c.}, 
 \label{eq:Yukawa}
\end{align}
where $\Phi^0$ and $\Phi^{\prime 0}$ are the neutral component of the Higgs doublets defined in Eq.~(\ref{eq:phi}). 
Here, we omitted the flavor indices. 
In the above expression, the dashed fermion fields denote their mass eigenstates, 
and $m_f$ ($f=u,d,\ell$) are the diagonalized mass matrices for charged fermions. 
The matrix $V$ represents the Cabbibo-Kobayashi-Maskawa matrix. 
It is seen that both the quark Yukawa interactions for $\Phi$ and $\Phi'$ are proportional to the diagonal matrices $m_u$ and $m_d$. 
Thus, the quark sector does not have FCNCs at tree level. 
On the other hand, the lepton sector contains the matrices defined as 
\begin{align}
M_\ell &= \frac{v}{\sqrt{2}}(c_\beta \tilde{Y}_\ell^1 + s_\beta \tilde{Y}_\ell^2),~~
Y_\ell^0 = U_L^\dagger \tilde{Y}_\ell U_R, \ Y_e = \tilde{Y}_\ell U_R, \notag\\
&\text{with}~~\tilde{Y}_\ell = -s_\beta \tilde{Y}_\ell^1 + c_\beta \tilde{Y}_\ell^2, \label{ye}
\end{align}
where $U_L$ and $U_R$ are respectively the unitary rotation matrices for $L_L^{}$ and $\ell_R^{}$. 
These unitary matrices diagonalize $M_\ell$ as 
\begin{align}
U_L^\dagger M_\ell U_R = m_\ell . \label{diag}
\end{align}
Since $\tilde{Y}_\ell$ is generally off-diagonal, we obtain the additional source of the lepton flavor violation. 
We note that the matrix elements of $\tilde{Y}_\ell$ can be written in terms of those for $M_\ell$, so that 
the structure of $\tilde{Y}_\ell$ is constrained such that the masses of charged leptons are reproduced. 
This is not the case in models with the Type-III Yukawa interaction, because both $\tilde{Y}_\ell^1$ and $\tilde{Y}_\ell^2$ are general complex $3\times 3$ matrices. 
As a consequence, the masses for charged leptons and neutrinos can be treated independently. 

\subsection{Higgs potential\label{sec:potential}}

The most general Higgs potential can be separately written by the following three parts:
\begin{align}
V = V_{\text{THDM}} + V_S + V_{\text{int}}, 
\end{align}
where $V_{\text{THDM}}$, $V_S$ and $V_{\text{int}}$ are functions of $(\Phi_1,\Phi_2)$, $S^\pm$ and $(\Phi_1,\Phi_2,S^\pm)$, respectively. 
Their explicit forms are given as 
\begin{align}
V_{\text{THDM}} &= m_1^2|\Phi_1|^2+m_2^2|\Phi_2|^2 -m_3^2(\Phi_1^\dagger\Phi_2 +  \text{h.c.})\notag\\
& +\frac{\lambda_1}{2}|\Phi_1|^4 + \frac{\lambda_2}{2}|\Phi_2|^4+\lambda_3|\Phi_1|^2|\Phi_2|^2+\lambda_4|\Phi_1^\dagger\Phi_2|^2,  \\
V_{S} &= m_S^2|S^+|^2
+\sigma_1|S^+|^2|\Phi_1|^2+\sigma_2|S^+|^2|\Phi_2|^2+\frac{\sigma_3}{2}|S^+|^4, \\
V_{\text{int}} & = \mu [\Phi_1^T (i\tau_2)\Phi_2 (S^+)^* + \text{h.c.}],
\end{align}
where phases of the $m_3^2$ and $\mu$ parameters can be absorbed by the phase redefinition of the scalar fields. 
These terms explicitly break the global $U(1)'$ symmetry. 
When we assign the lepton number of $-2$ ($+1$) unit for $S^+$ ($L_L$ and $\ell_R{}$) and zero for all the other fields, 
then the $\mu$ term explicitly breaks the lepton number with 2 units. 
This becomes the source of the Majorana neutrino mass term as it will be discussed in Sec.~\ref{sec:numass}. 

After solving the tadpole conditions for CP-even scalar bosons $h_1'$ and $h_2'$ defined by Eq.~(\ref{eq:phi}), 
we obtain the mass matrices for two CP-even and two charged scalar bosons as well as the mass of the CP-odd Higgs boson $A$. 
Their explicit forms are given in Appendix~\ref{sec:mass}. 
The stability of the Higgs potential has been studied in Ref.~\cite{Kanemura:2000bq}.  

\section{Predictions for the Lepton Sector\label{sec:numass}}

\begin{figure}[t]
\begin{center}
\includegraphics[width=150mm]{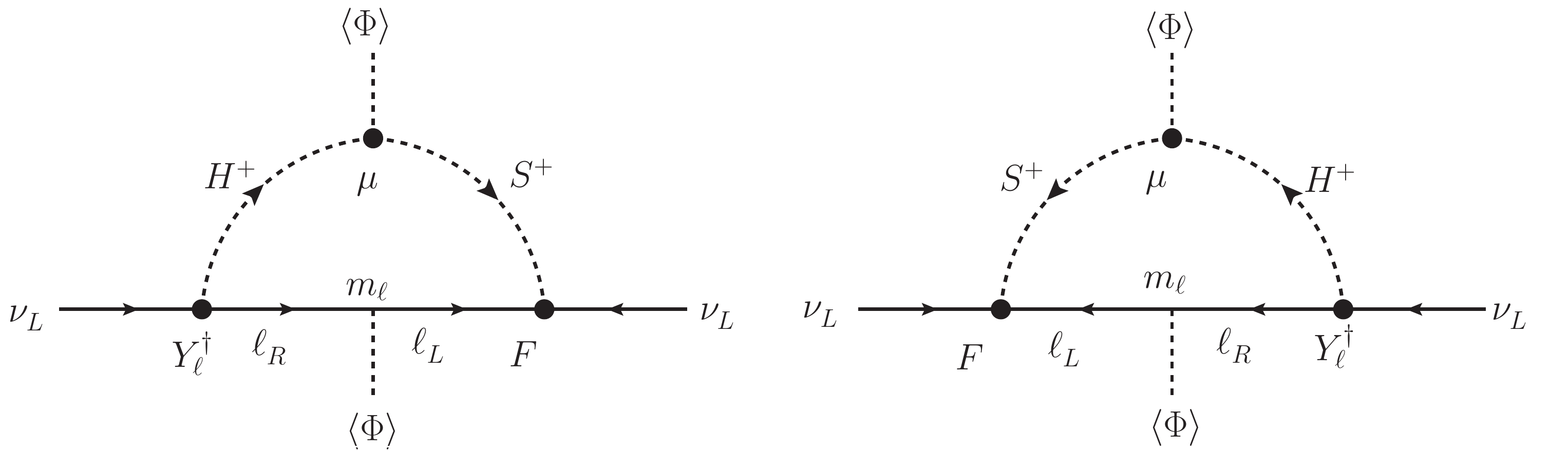}
\caption{One-loop diagram for the neutrino mass generation. The scalar bosons in this diagram are written 
in the Higgs basis. }
\label{fig1}
\end{center}
\end{figure}

\subsection{Neutrino masses and mixings}

Majorana masses for left-handed neutrinos are generated from the one-loop diagram as depicted in Fig.~\ref{fig1}. 
This diagram is calculated as follows: 
\begin{align}
  \mathcal{M}_\nu^{ij} &= C_\nu \big(F m_\ell Y_\ell^\dagger\big)^{ij} + (i\leftrightarrow j), \label{mnu}
\end{align}
where $F = \tilde{F}U_R$, and $C_\nu$ is the overall factor given as 
\begin{align}
C_\nu = \frac{1}{16\pi^2}\frac{\sqrt{2}v\mu}{m_{H_2^\pm}^2 - m_{H_1^\pm}^2} \ln \frac{m_{H_2^\pm}^2}{m_{H_1^\pm}^2} = \frac{s_{2\chi}}{16\pi^2} \ln \frac{m_{H_2^\pm}^2}{m_{H_1^\pm}^2}. \label{cnu}
\end{align}
The mass matrix given in Eq.~(\ref{mnu}) can be diagonalized by the Pontecorvo-Maki-Nakagawa-Sakata matrix $U_{\rm PMNS}$ as 
\begin{align}
U_{\rm PMNS}^T \, \mathcal{M_\nu} \, U_{\rm PMNS}  = \text{diag}(m_1,m_2,m_3), \label{mns}
\end{align}
where $m_i$ $(i=1,2,3)$ are the mass eigenvalues of the neutrinos. 
We note that the matrix $Y_\ell$ in Eq.~(\ref{mnu}) becomes $\sqrt{2}m_\ell \cot\beta/v$ when we consider the original Zee model, and the expression is consistent with that given in~\cite{He:2003ih}. 
As we already mentioned in Sec.~\ref{sec:yukawa}, 
the elements of $\tilde{Y}_\ell^{1,2}$ are constrained so as to reproduce the charged lepton masses. 
Therefore, the masses of charged leptons and neutrinos should be taken into account simultaneously. 
Namely, we cannot separately consider these two observables. 

Here, let us explain our strategy for the calculation of the masses of neutrinos and charged leptons. 
For simplicity, we consider the case without CP phases in the Yukawa interactions. 
First, from Eqs.~(\ref{ye}) and (\ref{diag}) the mass matrix $M_\ell $ is rewritten as 
\begin{align}
M_\ell  = \frac{v}{\sqrt{2}}(c_\beta \tilde{Y}_\ell^1 + s_\beta \tilde{Y}_\ell^2) = U_L m_\ell  U_R^T ,\label{me}
\end{align}
where $U_L$ and $U_R$ are now the $3\times 3$ orthogonal matrices:
\begin{align}
U_{L,R} = \begin{pmatrix}
1 &0&0 \\
0 & \cos\theta_{L,R}^{23} & -\sin\theta_{L,R}^{23} \\
0 & \sin\theta_{L,R}^{23} &  \cos\theta_{L,R}^{23}\\
\end{pmatrix}
\begin{pmatrix}
\cos\theta_{L,R}^{13} & 0 &-\sin\theta_{L,R}^{13} \\
0&1&0\\
\sin\theta_{L,R}^{13} & 0& \cos\theta_{L,R}^{13}\\
\end{pmatrix}
\begin{pmatrix}
\cos\theta_{L,R}^{12} & -\sin\theta_{L,R}^{12} &0\\
\sin\theta_{L,R}^{12} &  \cos\theta_{L,R}^{12} &0\\
0&0&1 \\
\end{pmatrix}. \label{ulr}
\end{align} 
From Eqs.~(\ref{me}) and (\ref{ulr}), each element of $\tilde{Y}_\ell^{1,2}$ is expressed in terms of the six angles ($\theta_{L,R}^{ij}$) and the charged lepton masses ($m_{\ell^i}$). 
We then can determine the matrix $Y_\ell$ from Eq.~(\ref{ye}). 
Finally, using Eqs.~(\ref{mnu})--(\ref{mns}) we obtain the predictions of three neutrino mixing angles $\theta_{12}$, $\theta_{23}$ and $\theta_{13}$
and three mass eigenvalues $m_i$. 
Clearly, the prediction for these neutrino observables depends on the choice of the charge assignment of the global $U(1)'$ symmetry, because it determines the structure of matrices 
$\tilde{Y}_\ell^{1,2}$ and $\tilde{F}$, see Appendix~\ref{sec:class}.

For the numerical evaluation of the neutrino mass matrix, we require that the predicted neutrino mixing angles $\theta_{ij}$ and the squared mass differences $\Delta m_{21}^2 \equiv m_2^2 - m_1^2$ and 
$\Delta m_{31}^2 \equiv m_3^2 - m_1^2$ are within the 2$\sigma$ range of the current experimental data given in \cite{deSalas:2017kay}.
In Class I of the $U(1)'$ charge assignment defined in Appendix~\ref{sec:class},  we take the following 11 independent parameters as inputs 
\begin{align}
\theta_{L,R}^{12},~
\theta_{L,R}^{23},~
\theta_{L,R}^{13},~
F_{12},~F_{23},~F_{13},~
\tan\beta,~C_\nu.  \label{para}
\end{align}
For Class II and Class III, one or two more matrix elements of $\tilde{F}$ becomes zero with respect to Class I, so that we cannot take all the elements of 
$F_{ij}$ as independent parameters. We will give further comments on these classes later. 
Among the parameters shown in Eq.~(\ref{para}), we fix $C_\nu$ defined in Eq.~(\ref{cnu}) to reproduce the best fit value of $\Delta m_{31}^2$ for each given set of input parameters.
Then, we scan the remaining 10 parameters, and see $\theta_{ij}$ and $\Delta m_{21}^2$ to be predicted within the 2$\sigma$ range. 

We find that only Class I can explain the current neutrino data at 2$\sigma$ level.
In Class II, no solution to satisfy all the neutrino data can be obtained after the scan analysis, as one or two zero elements appear in the matrix $\tilde{F}$, see Sec.~\ref{sec:class2}. 
We also verify that even if we take full $3 \times 3$ matrix elements in $\tilde{F}$, i.e, three independent nonzero elements, we cannot obtain the solution. 
This could be understood by the following way. 
First, the neutrino mass matrix can be rewritten as 
\begin{align}
  \mathcal{M}_\nu^{ij} &= \left\{
\begin{array}{l}
C_\nu [FU_L  (\cot \beta\, M_\ell P_1 M_\ell^\dagger  + \cot \beta \,M_\ell P_2 M_\ell^\dagger \cot \beta  - \tan \beta \,M_\ell P_3 M_\ell^\dagger ) ]^{ij} + (i \leftrightarrow j) \\
C_\nu [FU_L  (\cot \beta\, M_\ell M_\ell^\dagger P_1   + \cot \beta \,M_\ell M_\ell^\dagger P_2   - \tan \beta \,M_\ell M_\ell^\dagger P_3  ) ]^{ij} + (i \leftrightarrow j)
\end{array}\right., 
\end{align}
where the upper (lower) equation corresponds to the expression in Class I (Class II). 
We see the crucial difference between these two classes in the inserting position of the $P_i$ matrices defined in Eq.~(\ref{projection}). 
Namely in Class II, $P_i$ is multiplied at the end of each term, so that each term is projected by $P_i$, i.e., the term with $P_i$ only provides nonzero elements of the $i$-th column. 
On the other hand in Class I, the matrices $P_i$ are inserted at the middle, so that 
each term is not projected by $P_i$ at the end. 
Therefore, in Class I, the $\tan\beta$ or $\cot\beta$ factor appears in a mixed way, while in Class II either the $\tan\beta$ or $\cot\beta$ factor
appears in each column. 
This characteristic distribution of the $\tan\beta$ dependence might be disfavored to explain the neutrino data in Class II. 
Needless to say, Class III cannot explain the neutrino data as it can be regarded as the special case of Class II. 

From the above discussion, we take Class I of the $U(1)'$ charge assignment in what follows.

\subsection{Lepton flavor violations \label{sec:lfvv}}

Our Yukawa interactions given in Eq.~(\ref{eq:Yukawa}) induce charged lepton flavor violation (CLFV) processes\footnote{These also 
introduce flavor violating $Z$ boson decays at one-loop level. However, the size of the branching ratio is typically more than one order of magnitude smaller than the current upper limit~\cite{Koide:2000jm,Ghosal:2001ep}. }.
In the following, we discuss constraints from LFV processes in the alignment limit, i.e., $s_{\beta - \alpha} = 1$, in which 
all the couplings of the SM-like Higgs boson $h$ become the SM values at tree level. 
Thus, the LFV processes are induced via the extra Higgs bosons. 
Such configuration is also favored by the current LHC data~\cite{Sirunyan:2018koj,ATLAS:2018doi}.

We first consider $\ell_i \to \ell_j \gamma$ ($i > j$) processes, where $\{\ell_1,\ell_2,\ell_3\} = \{e, \mu, \tau\}$. 
Their branching ratios (BR) are calculated by neglecting the charged lepton mass in the final state as 
\begin{align}
&{\rm BR}(\ell_i\to\ell_j\gamma)\simeq \frac{48\pi^3\alpha_{\rm em}C_{ij}}{G_F^2 m^2_{\ell_i}} \left(\Big|\sum_\phi(a_R^\phi)_{ij}\Big|^2 + \Big|\sum_\phi(a_L^\phi)_{ij}\Big|^2\right), \label{br_clfv}
\end{align}
where $\alpha_{\rm em}$ and $G_F$ are the fine structure constant and the Fermi constant, respectively, and 
$C_{ij}$ are numerical constants given as $C_{21}=1$, $C_{31}=0.1784$ and $C_{32}=0.1736$. 
In Eq.~(\ref{br_clfv}), $a^\phi_{L,R}$ denote an amplitude obtained from a one-loop diagram with a scalar boson $\phi = \{H, A, H^\pm_1, H^\pm_2\}$ running in the loop. 
These amplitudes are explicitly given in the Appendix~\ref{sec:LFV_amp}.

Furthermore, the Yukawa interactions induce three body CLFV decays $\ell_i^\mp \to \ell_j^\mp \ell_k^\mp \ell_l^\pm$ at tree level by exchanging neutral scalar bosons.
Here, we focus on $\mu^\mp \to e^\mp e^\mp e^\pm$ and $\tau^\mp \to \mu^\mp \mu^\mp \mu^\pm$ processes.  
The other three body decays of $\tau$ are subdominant as compared to the $3\mu$ mode, 
because the couplings associated with the electron are included, which are significantly suppressed by $\mu \to e \gamma$.
The BRs of these processes are expressed by neglecting the mass of charged leptons as~\cite{Herrero-Garcia:2017xdu} 
\begin{align}
& {\rm BR}(\mu \to e e e) \simeq \frac{1}{64 G_F^2 m^4_{H}} \left[ |(Y_\ell^0)_{11}^* (Y_\ell^0)_{12} |^2 + |(Y_\ell^0)_{11} (Y_\ell^0)_{21}^* |^2 \right] {\rm BR}(\mu \to e \nu \nu), \\
&{\rm BR}(\tau \to \mu \mu \mu) \simeq \frac{1}{64 G_F^2 m^4_{H}} \left[ |(Y_\ell^0)_{22}^* (Y_\ell^0)_{23} |^2 + |(Y_\ell^0)_{22} (Y_\ell^0)_{32}^* |^2 \right] {\rm BR}(\tau \to \mu \nu \nu), 
\end{align}
where we have taken $m_{A} = m_{H}$.
There are also one-loop box diagram contributions to these processes with the charged Higgs bosons running in the loop.
However, their contributions are much smaller than the tree level one given in the above~\cite{Mituda:2001jw}, so that we can safely ignore such loop contributions.

We also consider a spin-independent $\mu \to e$ conversion via the $H$ exchange~\footnote{The CP-odd scalar boson $A$ exchange induces a spin-dependent $\mu \to e$ conversion process which is less constrained}.
The BR for the process is obtained such that~\cite{Kuno:1999jp,Kitano:2002mt,Davidson:2018kud}
\begin{align}
& {\rm BR}(\mu \to e) = \frac{32 G_F^2 m_\mu^5}{\Gamma_{\rm cap}} \left| C_{SL}^{pp} S^{(p)} + C_{SL}^{nn} S^{(n)} + C_{SR}^{pp} S^{(p)} + C_{SR}^{nn} S^{(n)} \right|^2, \\
& C_{SL[SR]}^{pp} \simeq C_{SL[SR]}^{nn} \simeq \frac{f_N m_N}{2 G_F m_H^2 v } \cot \beta (Y^0_\ell)_{12[21]},
\end{align}
where $S^{(p,n)}$ is the integral over the nucleus for lepton wave functions with the corresponding nucleon density, 
$\Gamma_{\rm cap}$ is the rate for the muon to transform to a neutrino by capture on the nucleus, 
and $f_N \sim 0.3$ is the effective coupling between a Higgs boson and a nucleon $N$ defined by $f_N m_N \bar N N = \sum_q m_q \langle N | \bar q q |  N \rangle $ with a nucleon mass $m_N$~\cite{Cline:2013gha}.
The values of $\Gamma_{\rm cap}$ and $S^{(n,p)}$ depend on target nucleus, and those for $^{197}_{79}$Au and $^{27}_{13}$Al targets  are given by 
$\Gamma_{\rm cap}(^{197}_{79}{\rm Au} [^{27}_{13}{\rm Al}] ) = 13.07[0.7054] \times 10^6~ {\rm sec}^{-1}$,  $S^{(p)}(^{197}_{79} {\rm Au} [^{27}_{13}{\rm Al}] ) = 0.0614[0.0155]$ and 
$S^{(n)}(^{197}_{79} {\rm Au} [^{27}_{13}{\rm Al}] ) = 0.0981[0.0167]$~\cite{Suzuki:1987jf,Kitano:2002mt}.

The current upper limits on the above BRs with 95\% confidence level are given in Refs.~\cite{TheMEG:2016wtm, Aubert:2009ag,Renga:2018fpd,Lindner:2016bgg}
for the $\ell_i \to \ell_j \gamma$ processes, in Refs.~\cite{Bellgardt:1987du, Hayasaka:2010np} for the $\mu \to 3e$ and $\tau \to 3\mu$ processes 
 and in Refs.~\cite{Bertl:2006up,Coy:2018bxr} for the $\mu \to e$ process:
\begin{align}
&{\rm BR}(\mu\to e\gamma) < 4.2\times10^{-13},~ 
{\rm BR}(\tau\to e\gamma) < 3.3\times10^{-8},~
{\rm BR}(\tau\to\mu\gamma) < 4.4\times10^{-8}, \nonumber  \\
&{\rm BR}(\mu \to e e e) < 1.0 \times 10^{-12},~ {\rm BR}(\tau \to \mu \mu \mu) < 2.1 \times 10^{-8}, \nonumber \\
&  {\rm BR}(\mu \to e)_{\rm Al} < 7 \times 10^{-13}. \label{eq:lfvs-cond}
\end{align}
We impose them in our numerical analysis below.

Before closing this section, let us calculate the decay rates of the additional neutral Higgs bosons into leptons, i.e., $H \to \ell^+_i \ell^-_j$ and $A \to \ell_i^+ \ell_j^-$. 
Because of the LFV couplings, the final state leptons can be either same flavor or different flavor, where the latter does not 
happen in the THDMs with a softly-broken $Z_2$ symmetry. 
The expressions for the decay rates are given in the alignment limit $s_{\beta - \alpha} =1$ by 
\begin{align}
& \Gamma(\phi^0 \to \ell_i \ell_j) = \frac{1}{32 \pi (1+ \delta_{ij})} m_{\phi^0} (|(Y^0_\ell)_{ij} |^2 + |(Y^0_\ell)_{ji}|^2), 
\end{align}
where $\ell^+_i \ell^-_j$ and $\ell_j^+ \ell_i^-$ modes for $i\neq j$ are summed. 
The decay rates for quark final states are the same as in the Type-I THDM.
We note that the decay modes of $H$ ($A$) into $ZZ/W^+W^-/hh$ ($Zh$) are absent in the alignment limit at tree level.

\section{Numerical results \label{sec:pheno}}

\begin{figure}[t!]\begin{center}
\includegraphics[width=60mm]{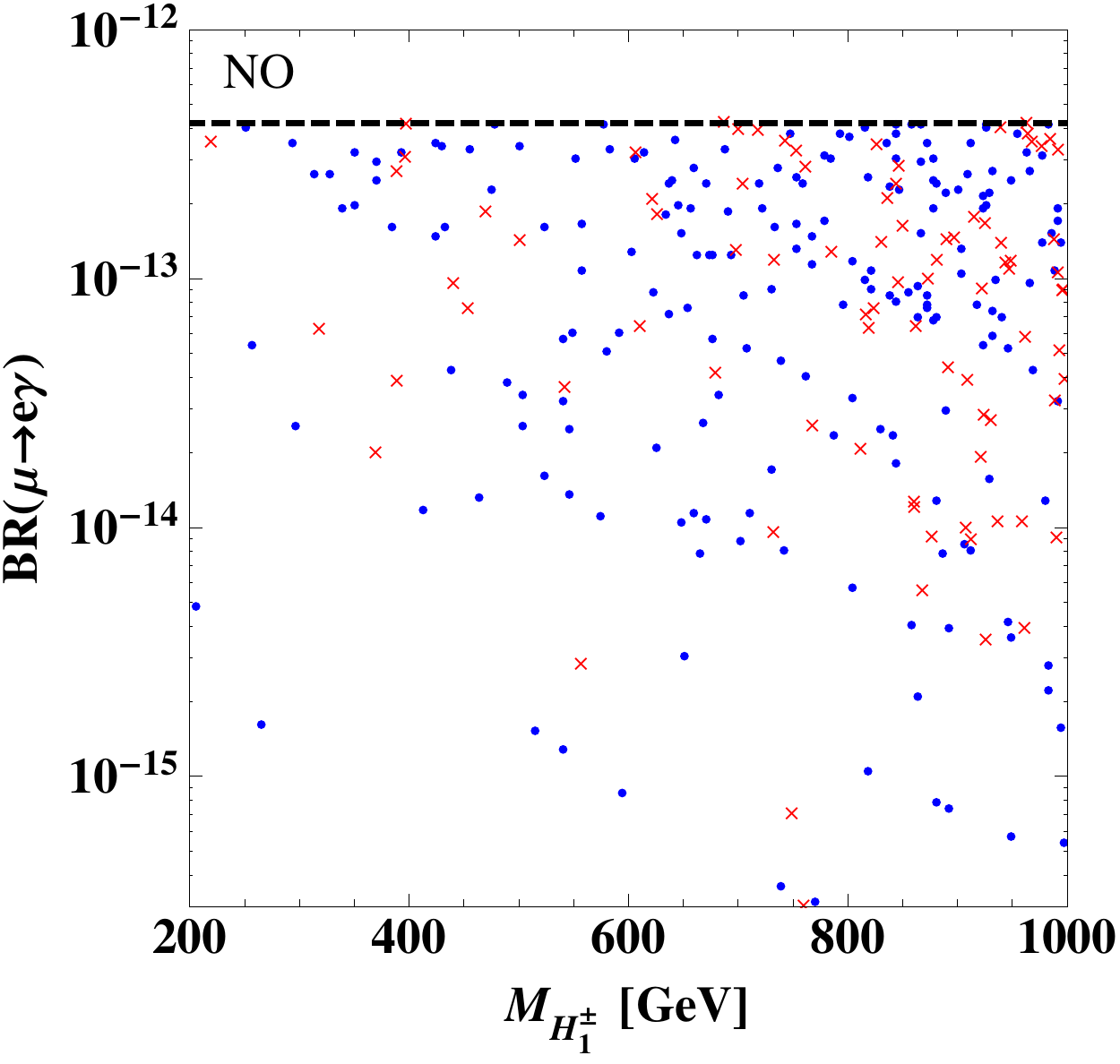} \qquad
\includegraphics[width=60mm]{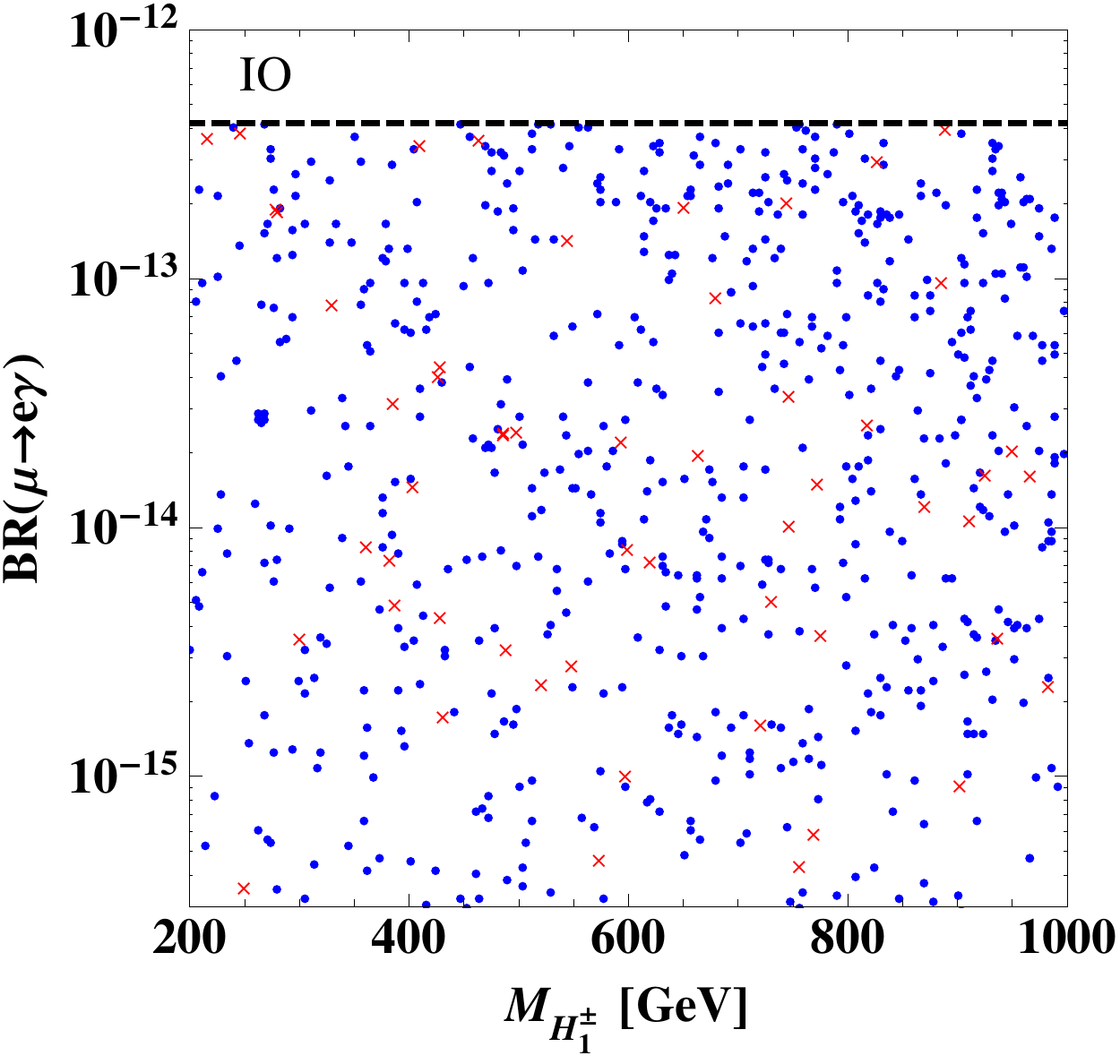} 
\includegraphics[width=60mm]{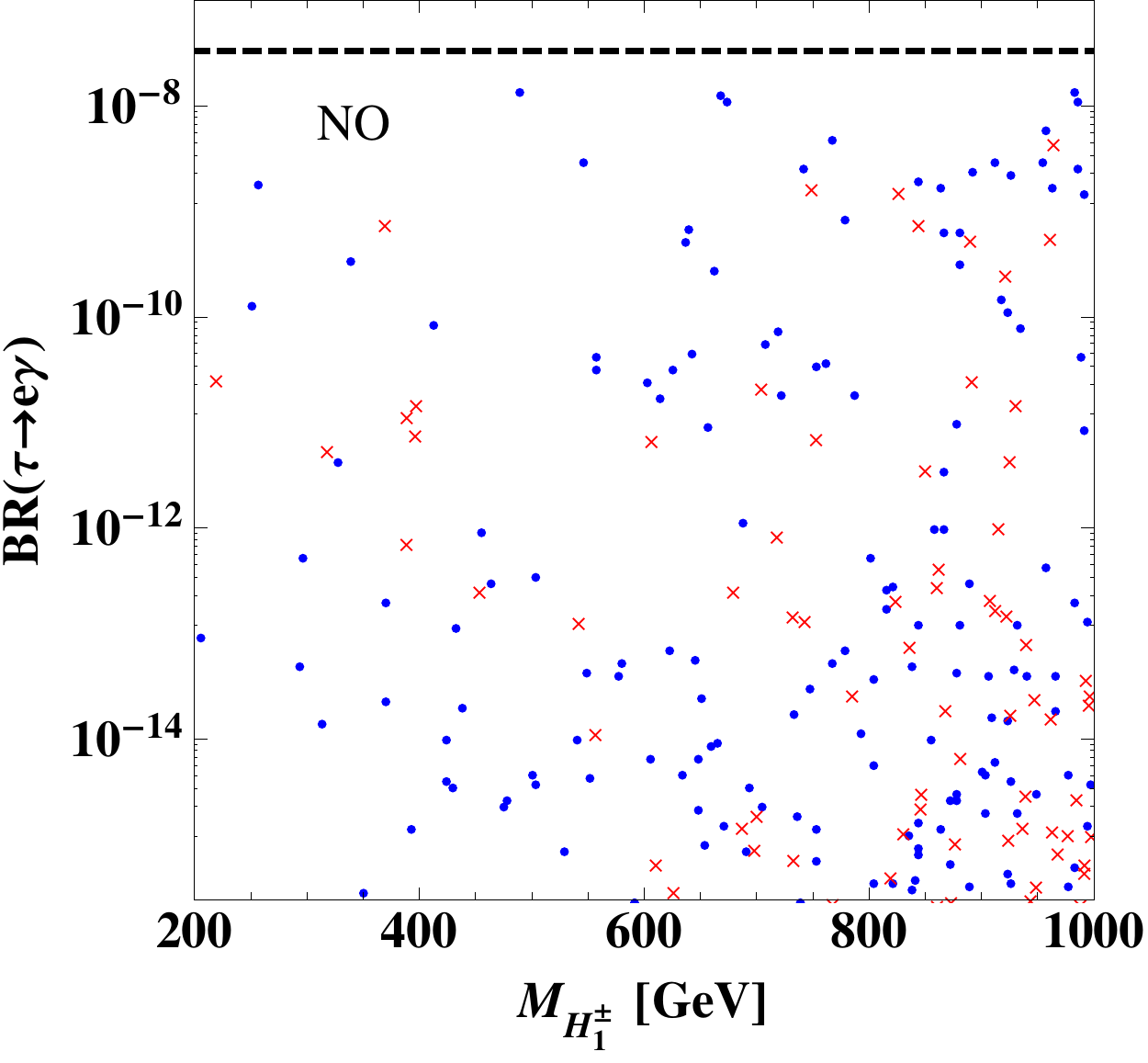} \qquad
\includegraphics[width=60mm]{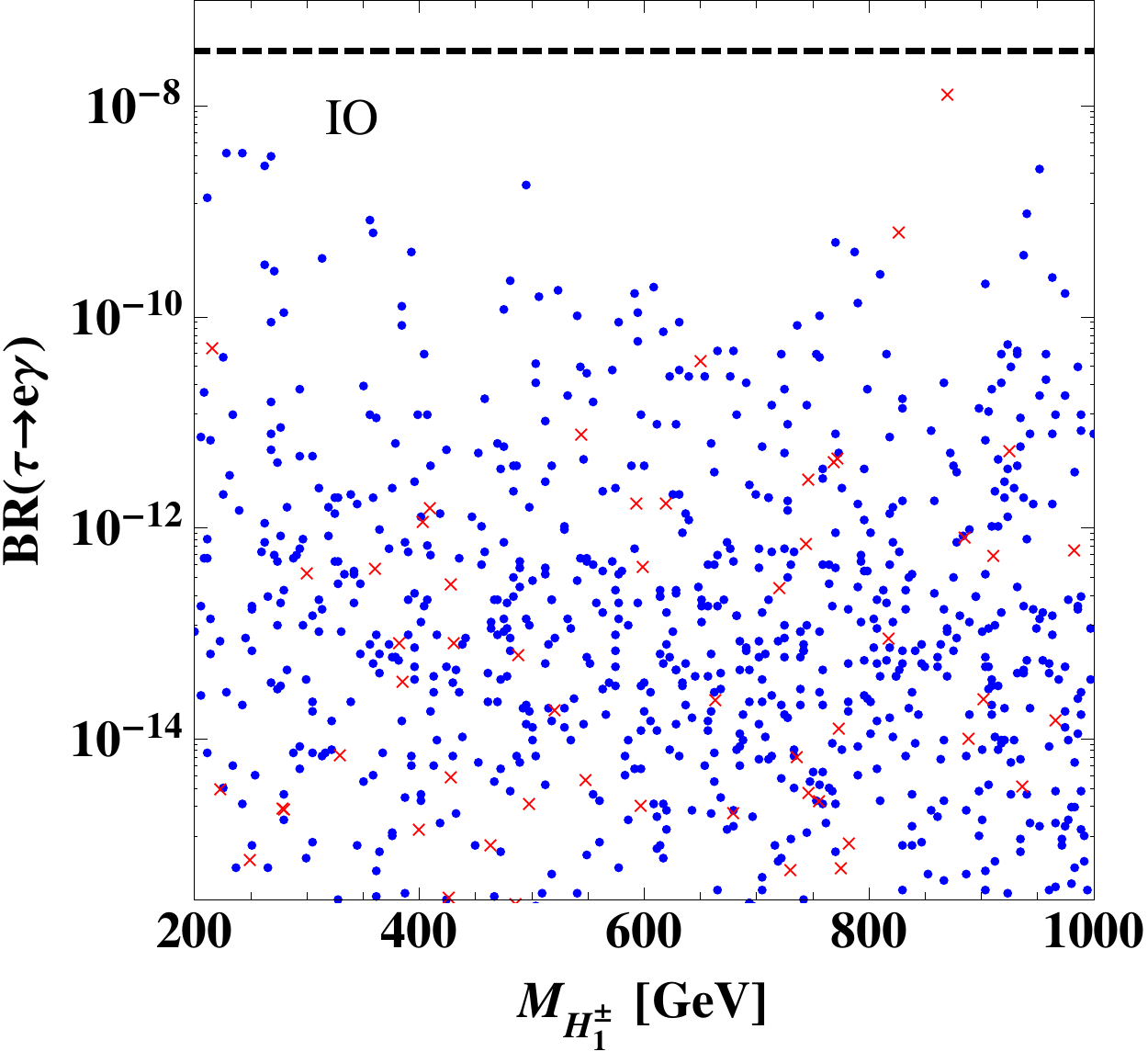}
\includegraphics[width=60mm]{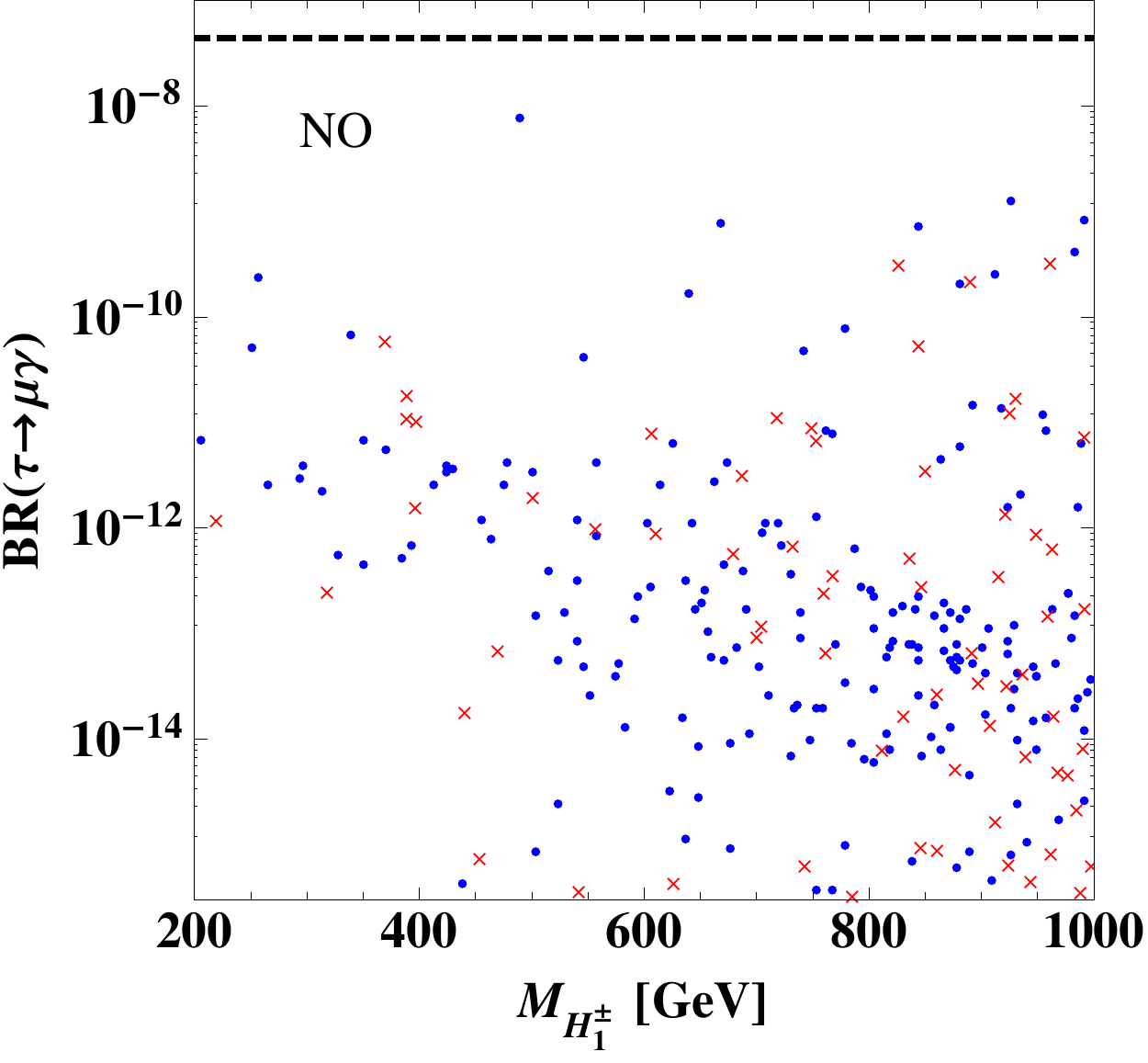} \qquad
\includegraphics[width=60mm]{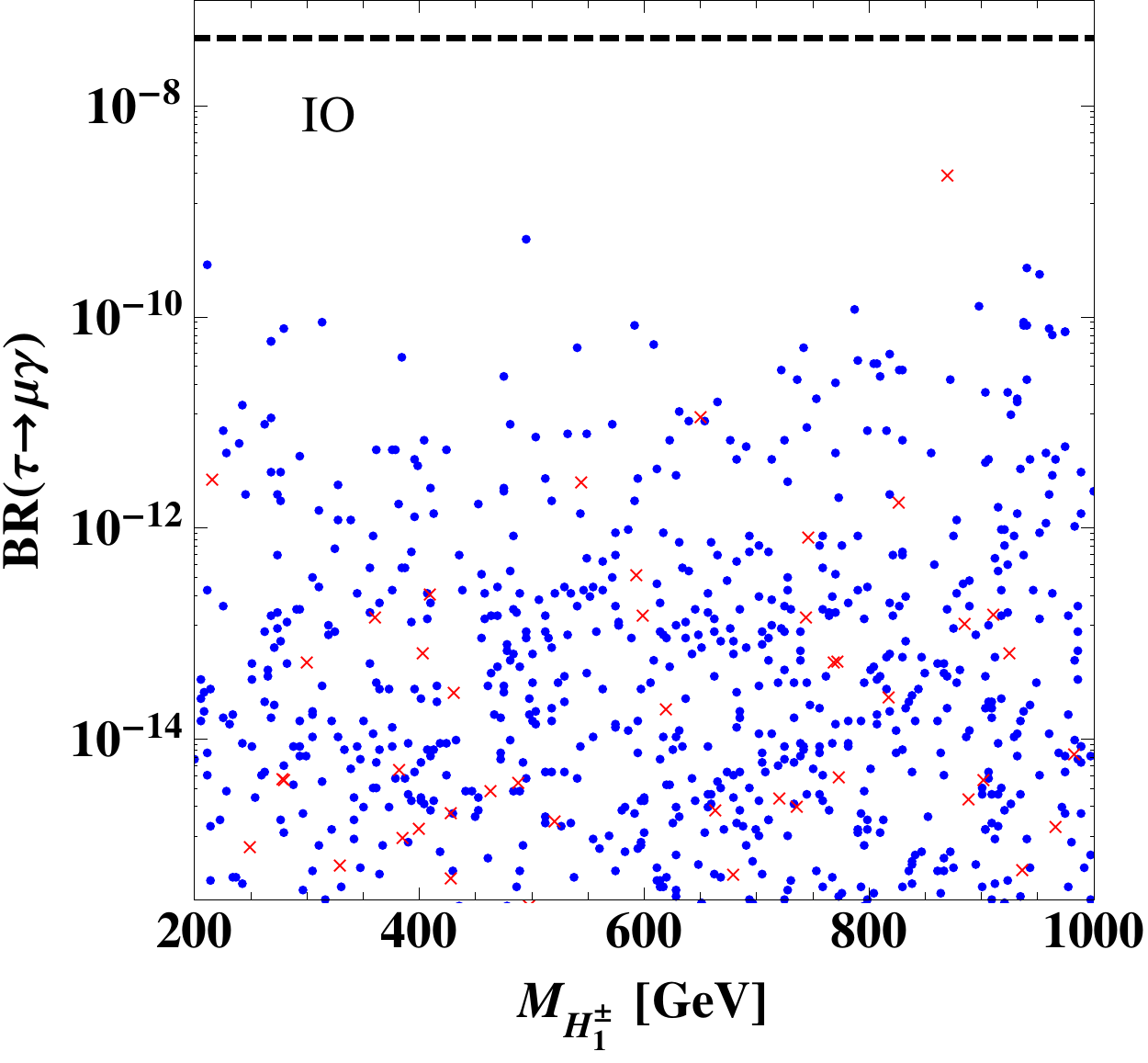}
\caption{BRs for the $\ell_i \to \ell_j \gamma$ processes as a function of $m_{H_1^\pm}$  in the NO case (left) and the IO case (right). 
The blue dots (red crosses) show the case with $1< \tan\beta <10$ $(10 < \tan \beta < 30)$.
The dashed horizontal line indicates the upper bounds for each BR with 95\% confidence level. }   
\label{fig:LFV1}
\end{center}
\end{figure}

\begin{figure}[t!]\begin{center}
\includegraphics[width=60mm]{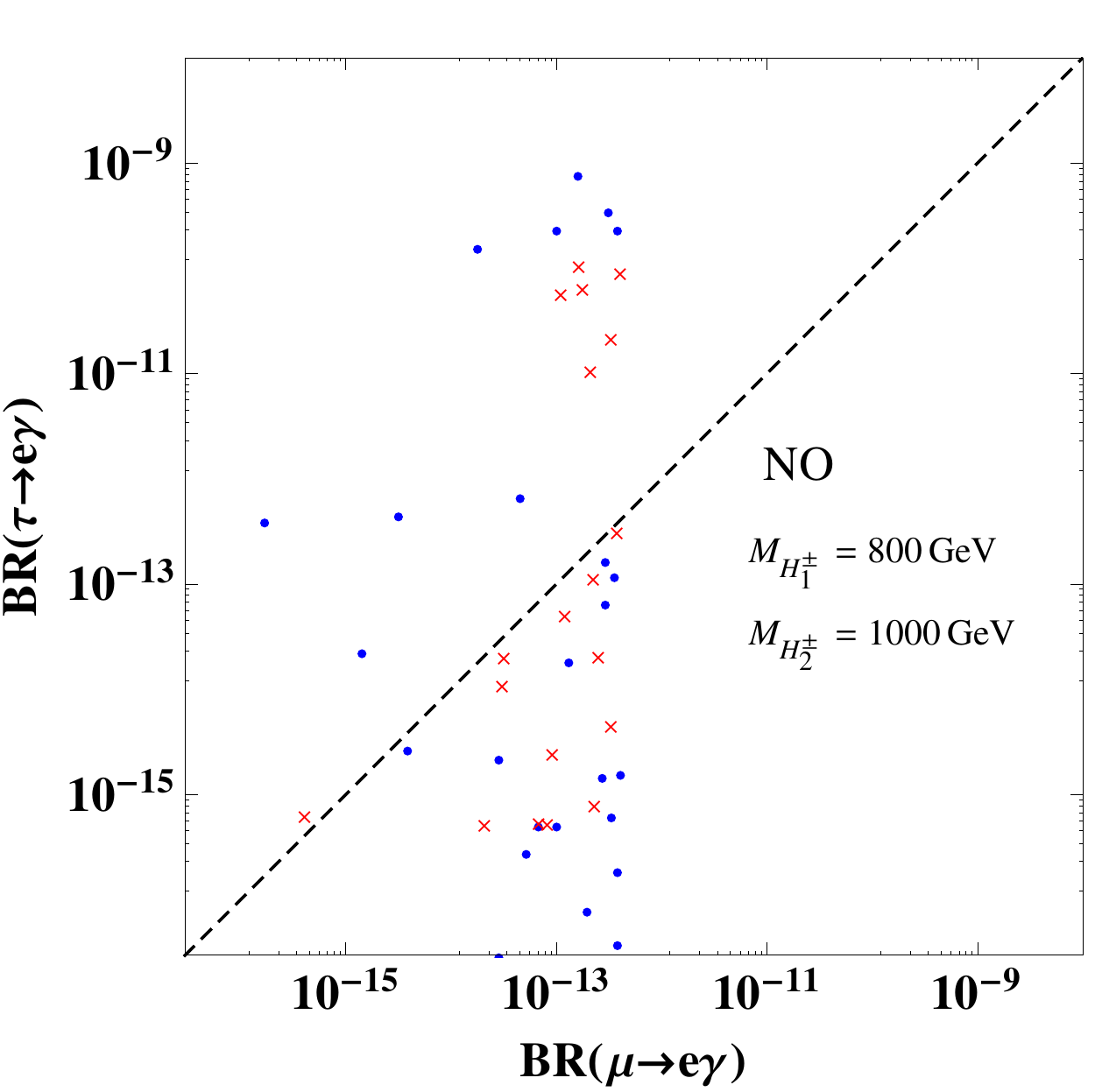} \qquad
\includegraphics[width=60mm]{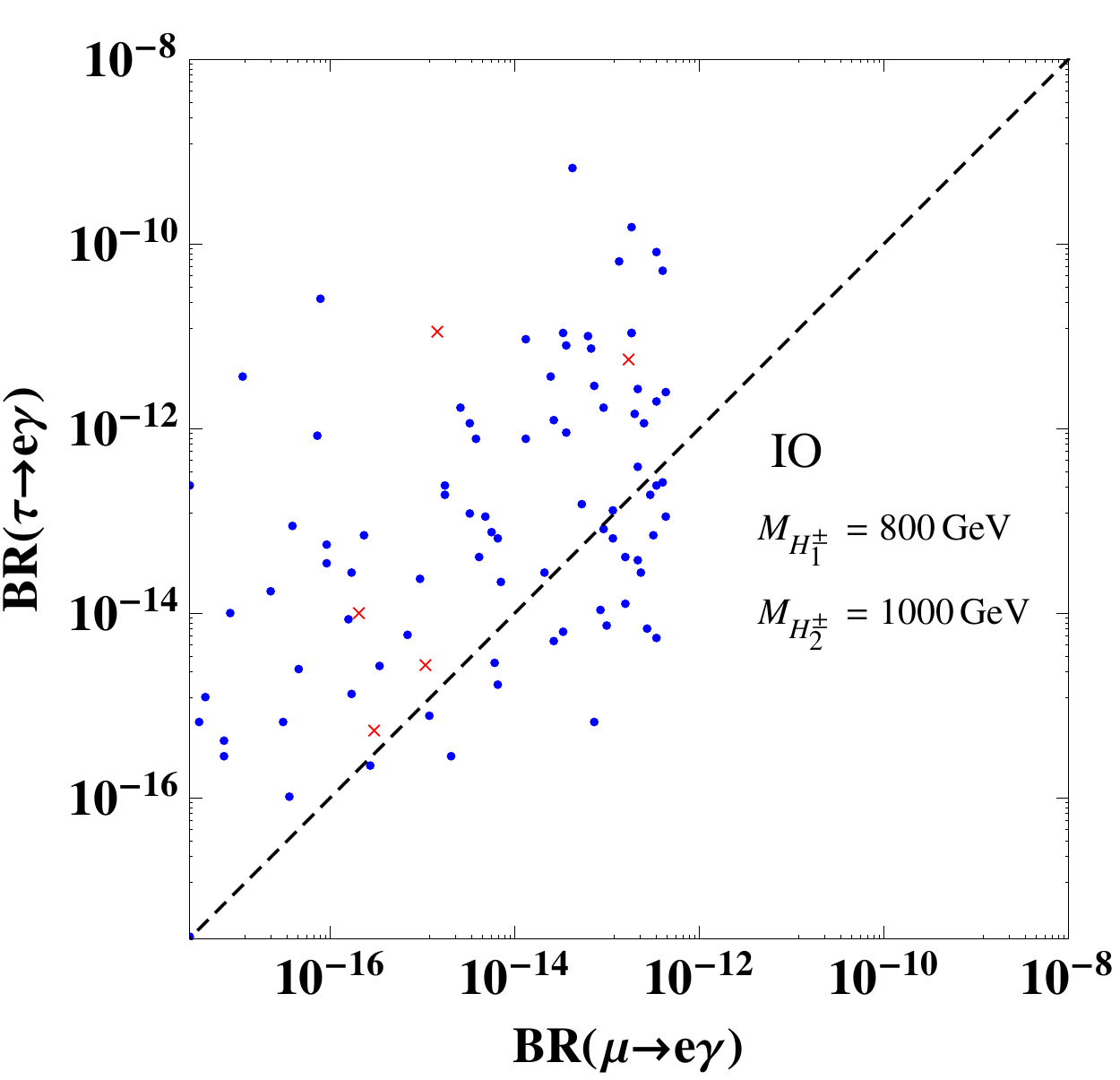}
\includegraphics[width=60mm]{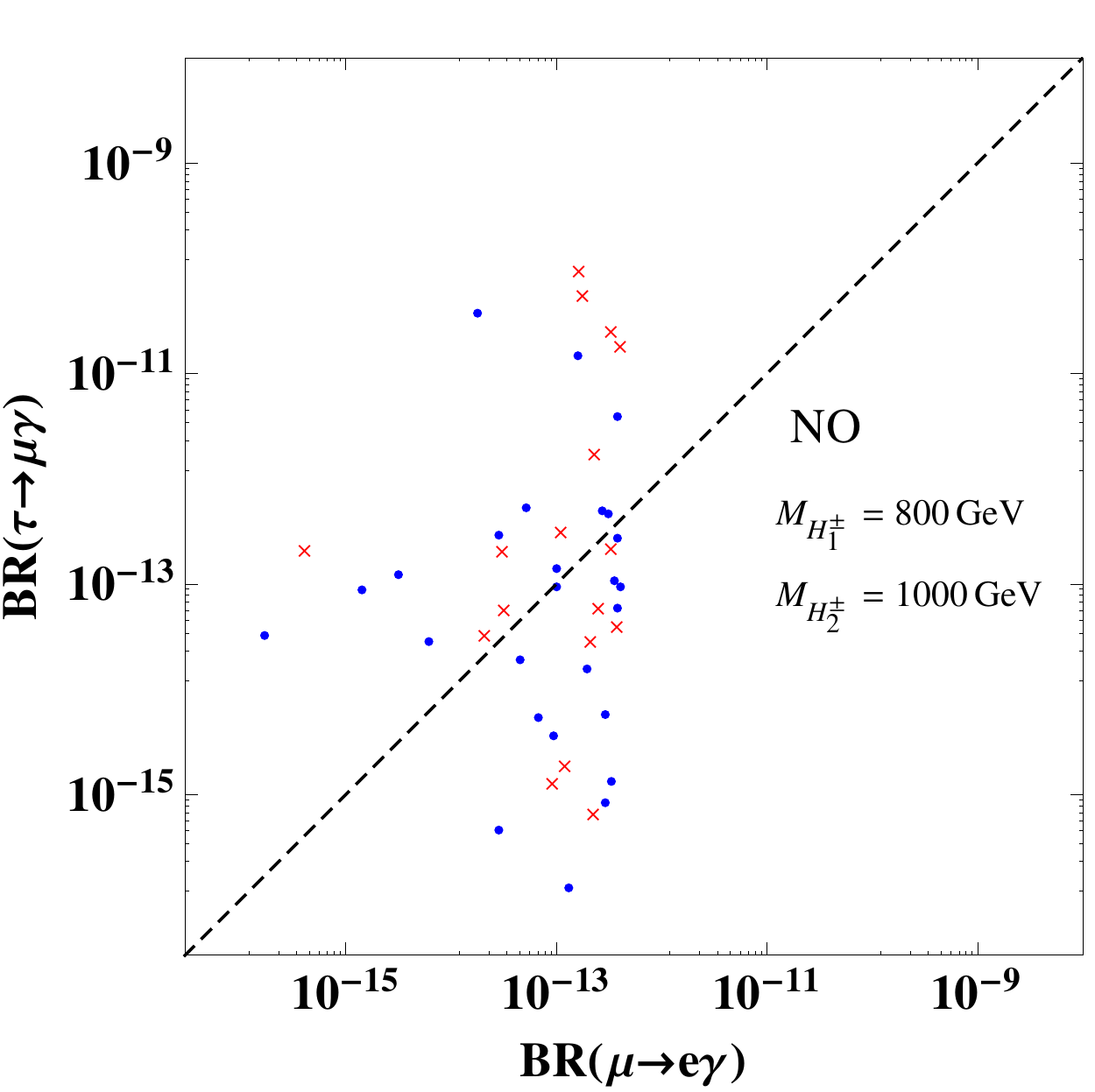} \qquad
\includegraphics[width=60mm]{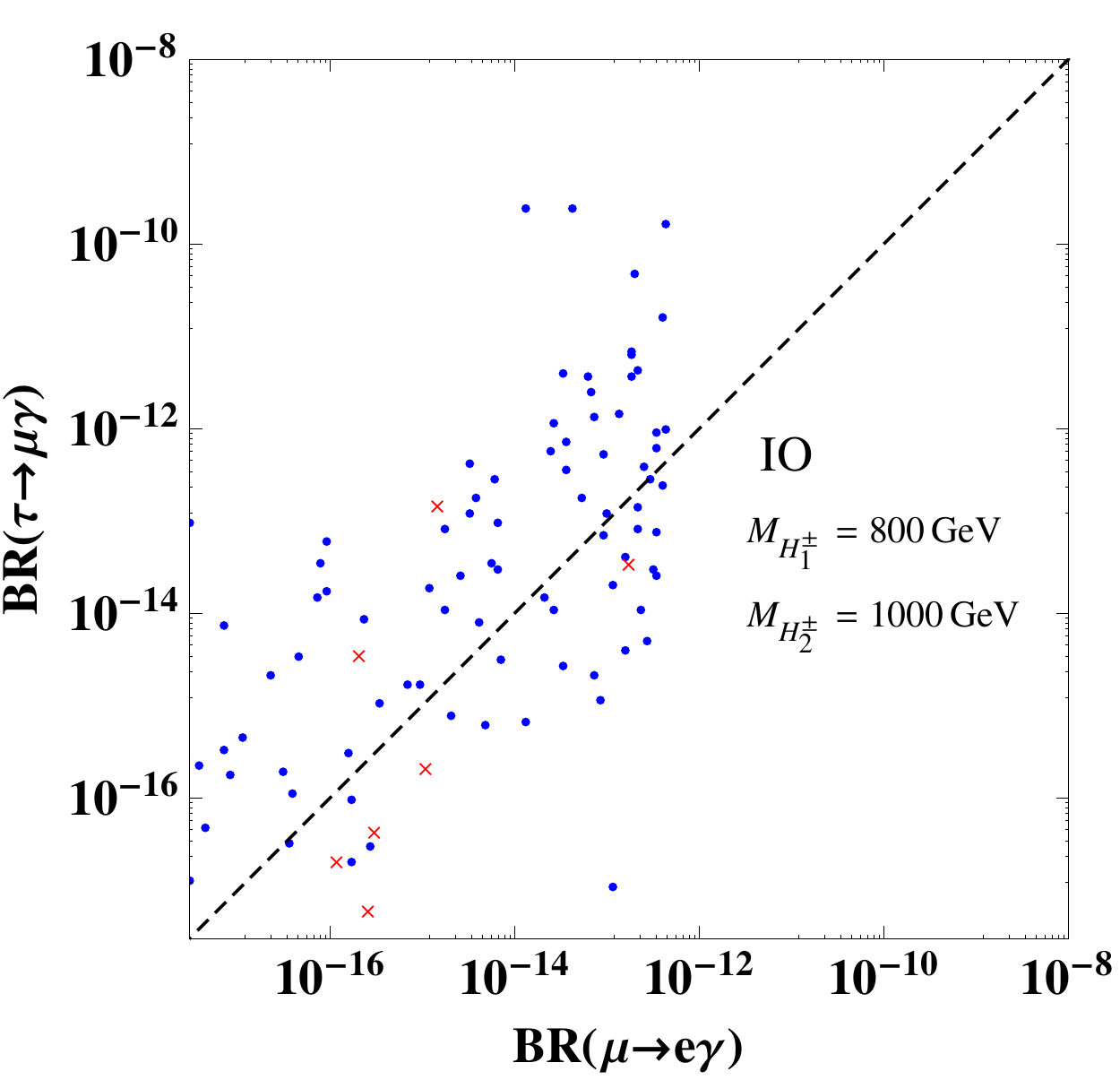}
\includegraphics[width=60mm]{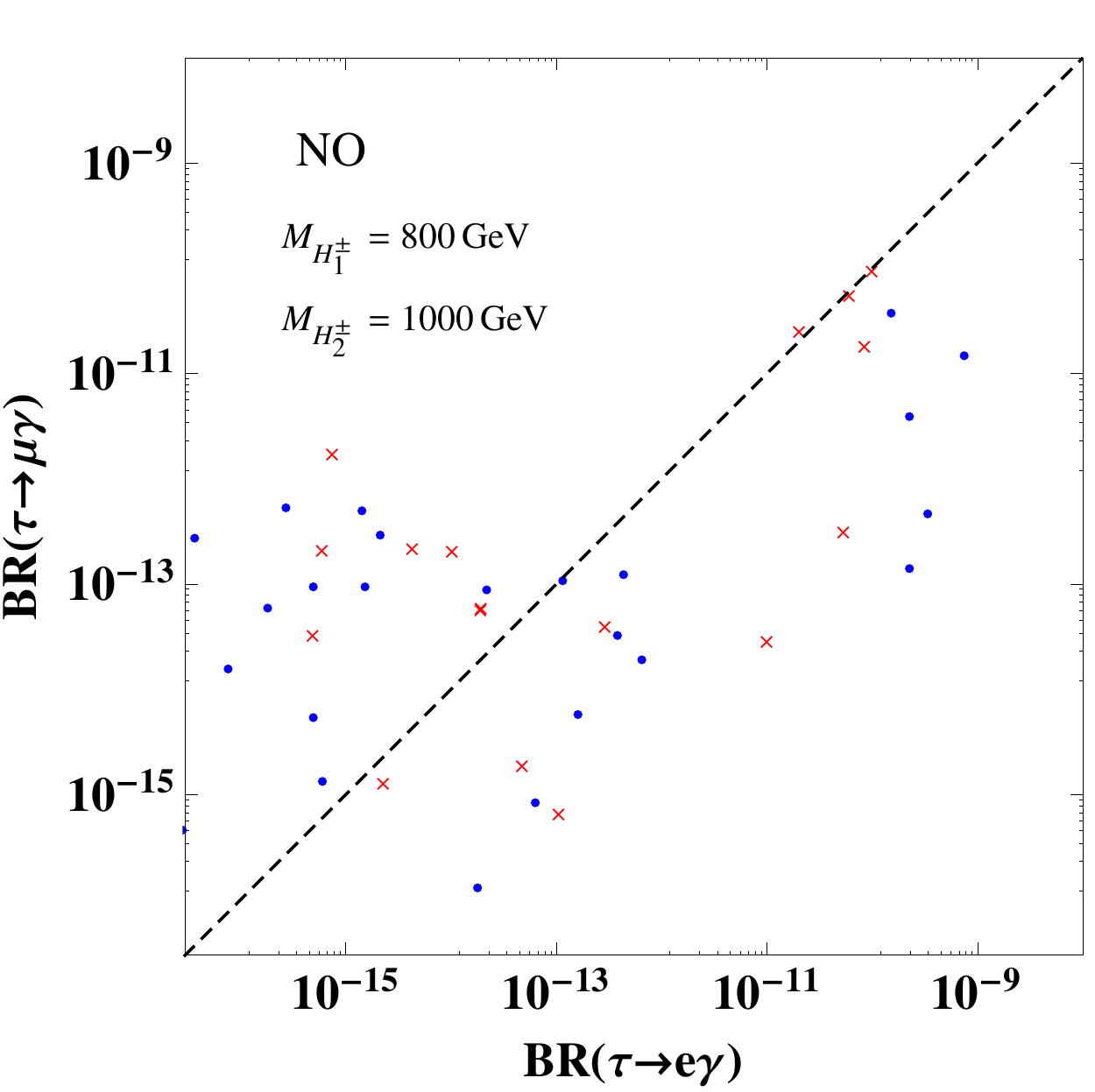} \qquad
\includegraphics[width=60mm]{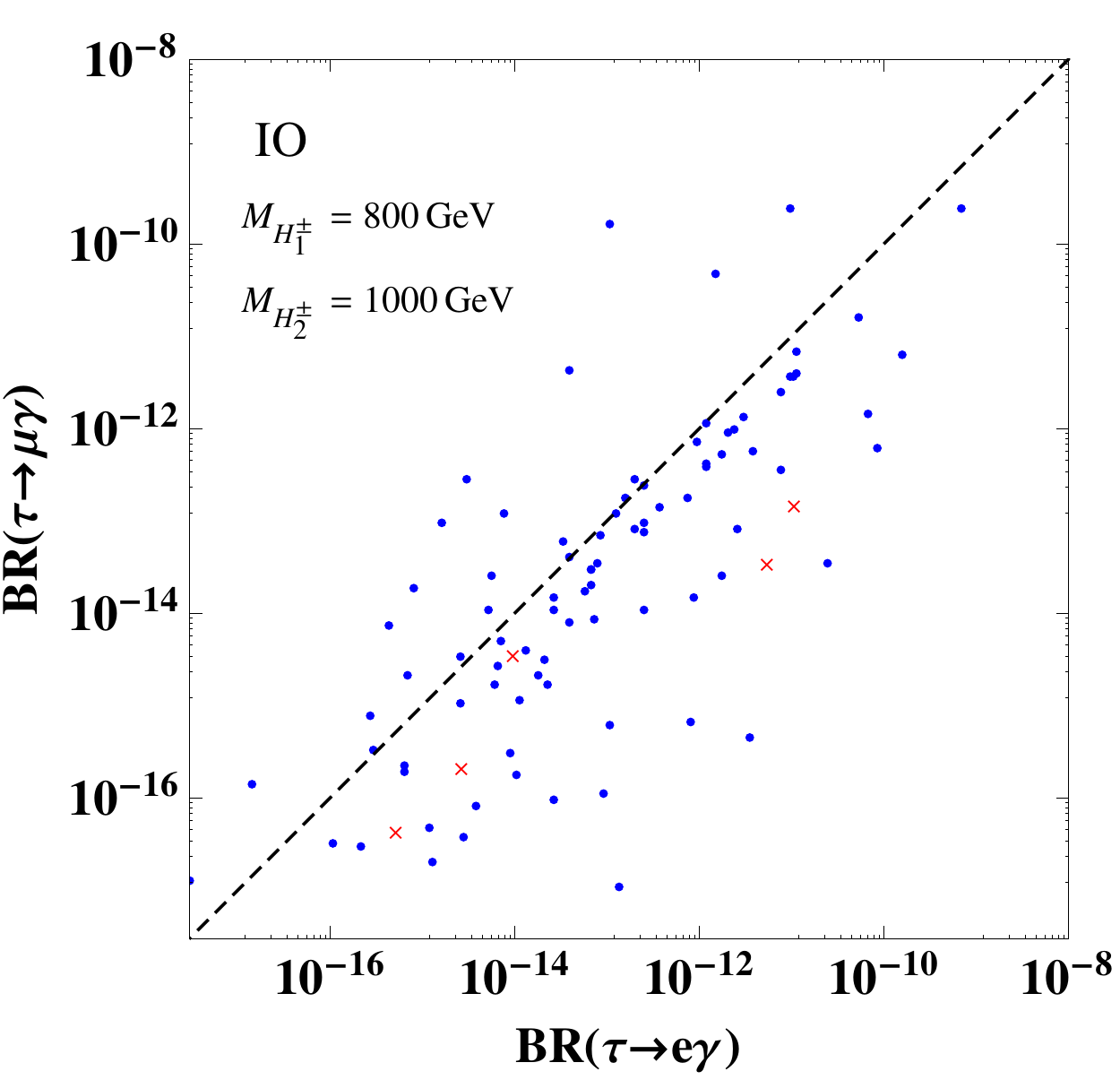}
\caption{Correlations among BR$(\ell_i \to \ell_j \gamma)$ with $m_{H_1^\pm}^{} = 800$ GeV and $m_{H_2^\pm}^{} = 1000$ GeV in the NO case (left) and the IO case (right). 
The blue dots (red crosses) show the case with $1< \tan\beta <10$ $(10 < \tan \beta < 30)$.
}   
\label{fig:LFV2}\end{center}\end{figure}

In this section, we numerically evaluate the BRs for the CLFV modes, i.e.,  $\ell_i \to \ell_j \gamma$, $\mu \to 3e$ and $\tau \to 3\mu$, and 
those for the additional neutral Higgs bosons, under which the model parameters accommodate with neutrino oscillation data.
As we already mentioned in Sec.~\ref{sec:lfvv}, we take the alignment limit $s_{\beta-\alpha} = 1$ to avoid the constraints from LHC. 
In addition, to avoid the constraints from the electroweak $S$ and $T$ parameters~\cite{Peskin:1990zt,Peskin:1991sw}, we take $m_H^{} = m_A^{} = m_{H_1^\pm}^{}$, by which 
new contributions to the $S$ and $T$ parameters almost vanish\footnote{Tiny contributions to the $S$ and $T$ parameters remain, which exactly vanish at the limit of $s_\chi \to 0$.}.

Our input parameters are given in Eq.~(\ref{para}) for the calculation of the neutrino masses. 
However, for the calculation of the other observables such as BRs for the additional Higgs bosons, 
it is better to choose the following parameters as inputs: 
\begin{align}
\theta_{L,R}^{12},~
\theta_{L,R}^{23},~
\theta_{L,R}^{13},~
F_{12},~F_{23},~F_{13},~
\tan\beta,~s_\chi,~m_{H_1^\pm}^{},~m_{H_2^\pm}^{},~m_{H}^{},~m_{A}^{},~s_{\beta-\alpha}.  \label{para2}
\end{align}
The first 10 parameters, except for the overall factor of $F_{ij}$, are determined such that the neutrino data and charged lepton masses are reproduced. 
The 3 parameters $s_\chi,~m_{H_1^\pm}^{}$ and $m_{H_2^\pm}^{}$ determine the overall factor of the neutrino mass matrix $C_\nu'$ (here let us denote it by $C_\nu'$, not $C_\nu$). 
The correct value of $C_\nu$ to reproduce the neutrino data is then obtained by multiplying $C_\nu'/C_\nu$ to $F_{ij}$. 
If we do not specify the values of $s_\chi,~m_{H_1^\pm}^{}$ and $m_{H_2^\pm}^{}$, these parameters are scanned in the following ranges: 
\begin{equation}
s_\chi \in [10^{-6}, 10^{-2}], \quad m_{H_1^\pm}^{} \in [200, 1000] \ {\rm GeV}, \quad m_{H_2^\pm}^{} \in [m_{H_1^\pm}^{}, 1500] \ {\rm GeV},
\end{equation}

In Fig.~\ref{fig:LFV1}, we show BR$(\ell_i \to \ell_j \gamma)$ as a function of $m_{H_1^\pm}^{}$ 
assuming the normal ordering (NO) case (left panel) and the inverted ordering (IO) case (right panel) for the neutrino mass hierarchy. 
We find that BR$(\mu \to e \gamma)$ tends to be smaller in the IO case as compared with the NO so that the former case is less constrained by the process.
In addition in the IO case, BR$(\tau \to e \gamma)$ and BR$(\tau \to \mu \gamma)$ tend to be suppressed for larger $\tan \beta$ case, because of the constraints from the neutrino oscillation data. 
We also find that the BRs of $\mu \to eee$ and $\tau \to \mu \mu \mu$ processes are typically one or two order of magnitude smaller than the current upper limit in the 
parameter sets allowed by the constraint from the $\ell_i \to \ell_j \gamma$ processes. 
It is due to the smallness of the diagonal elements $(Y^0_\ell)_{11,22}$ which is required from consistency with charged lepton mass and neutrino data.
We thus do not show the plots for three body LFV decay processes.
In addition, the maximal value of BR$(\mu \to e)_{\rm Au, Al}$ is around $\sim 10^{-13}$ for both the NO and IO cases when the BR($\mu \to e \gamma$) constraint is satisfied.
Therefore, it is safe from the current constraint in Eq.~(\ref{eq:lfvs-cond}), and we do not show corresponding scattering plots here.
In future experiments this process will be tested with high precision up to BR$ \sim 10^{-16}$~\cite{Kuno:2013mha, Carey:2008zz}, and the parameter space of our model can be further tested.

Furthermore, correlations between two of three BRs are shown in Fig.~\ref{fig:LFV2}.  
Here, we take the relatively larger scalar boson masses, i.e., 
$m_{H_1^\pm}^{} = 800$ GeV and $m_{H_2^\pm}^{} = 1000$ GeV to obtain more allowed parameter points, but the pattern of correlations does not change so much if we change the masses. 
We find that the BRs are more strongly correlated in the IO case as compared with the NO case, and the value of $\tan \beta$ does not much affect the correlation pattern.
In particular, we can see the tendency in the IO case that 
${\rm BR}(\tau\to e\gamma) \gtrsim {\rm BR}(\mu\to e\gamma)$, 
${\rm BR}(\tau\to \mu\gamma) \simeq {\rm BR}(\mu\to e\gamma)$ and 
${\rm BR}(\tau\to \mu\gamma) \lesssim {\rm BR}(\tau\to e\gamma)$. 

\begin{figure}[t!]\begin{center}
\includegraphics[width=60mm]{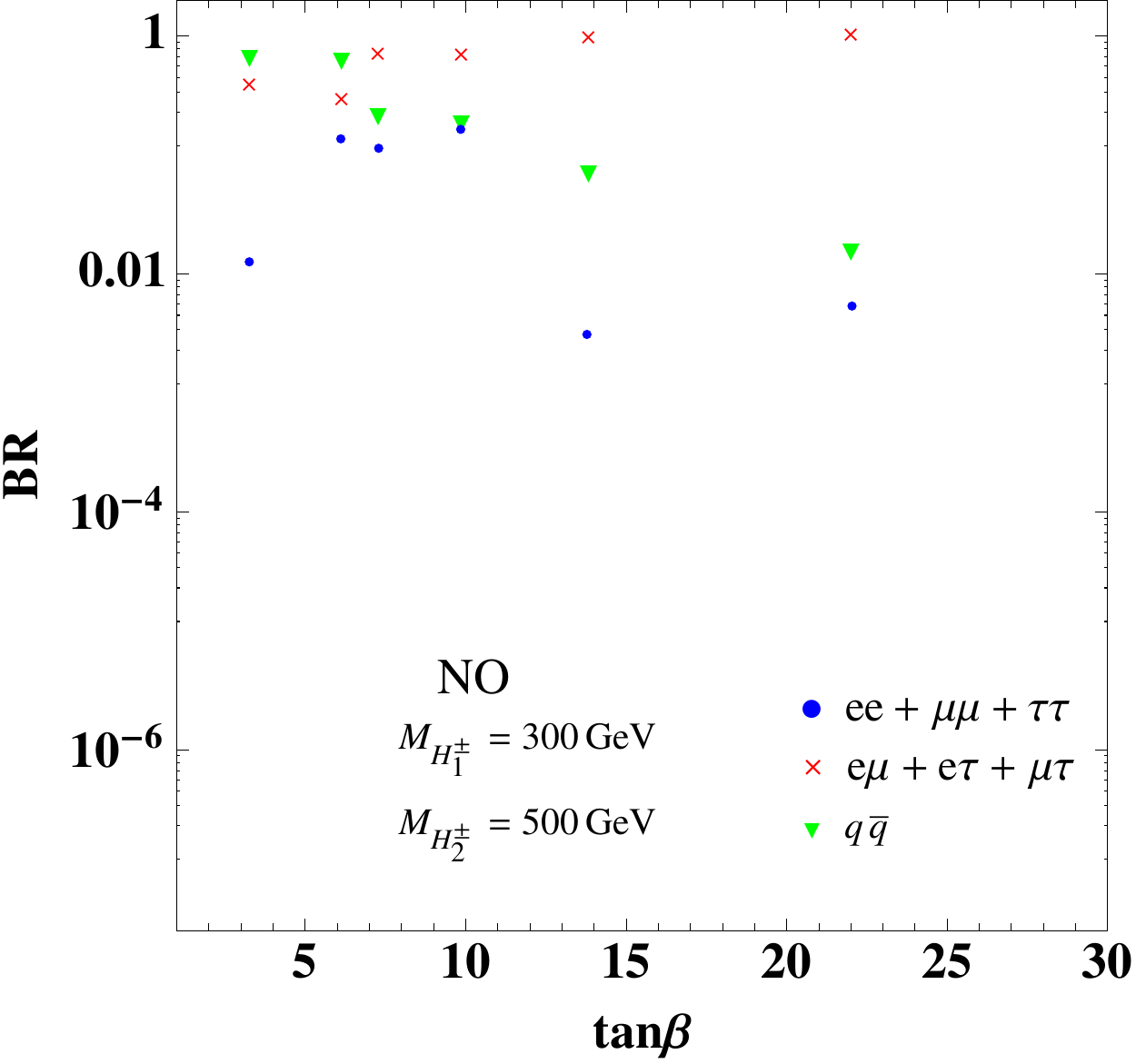} \qquad
\includegraphics[width=60mm]{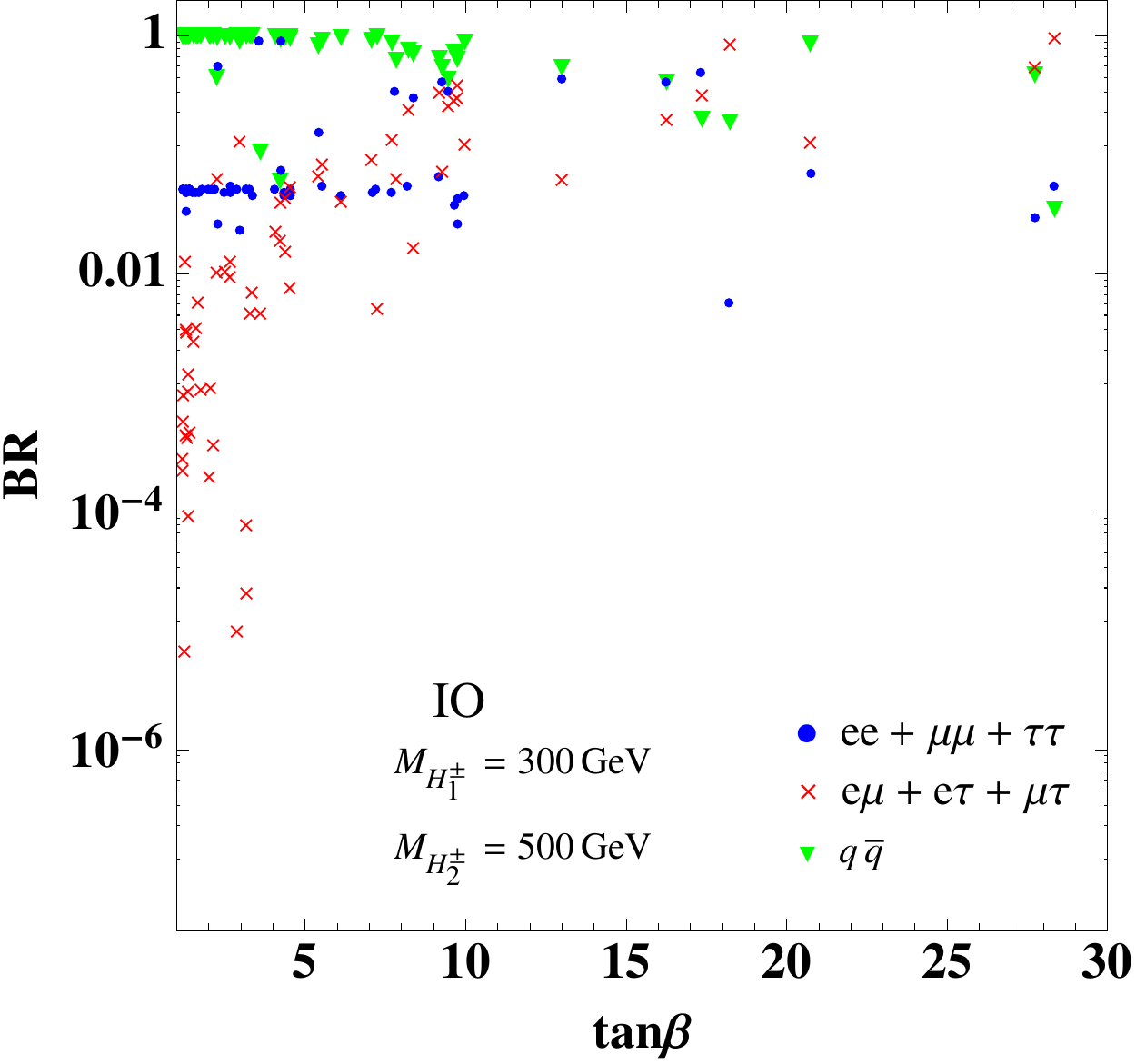} \\
\includegraphics[width=60mm]{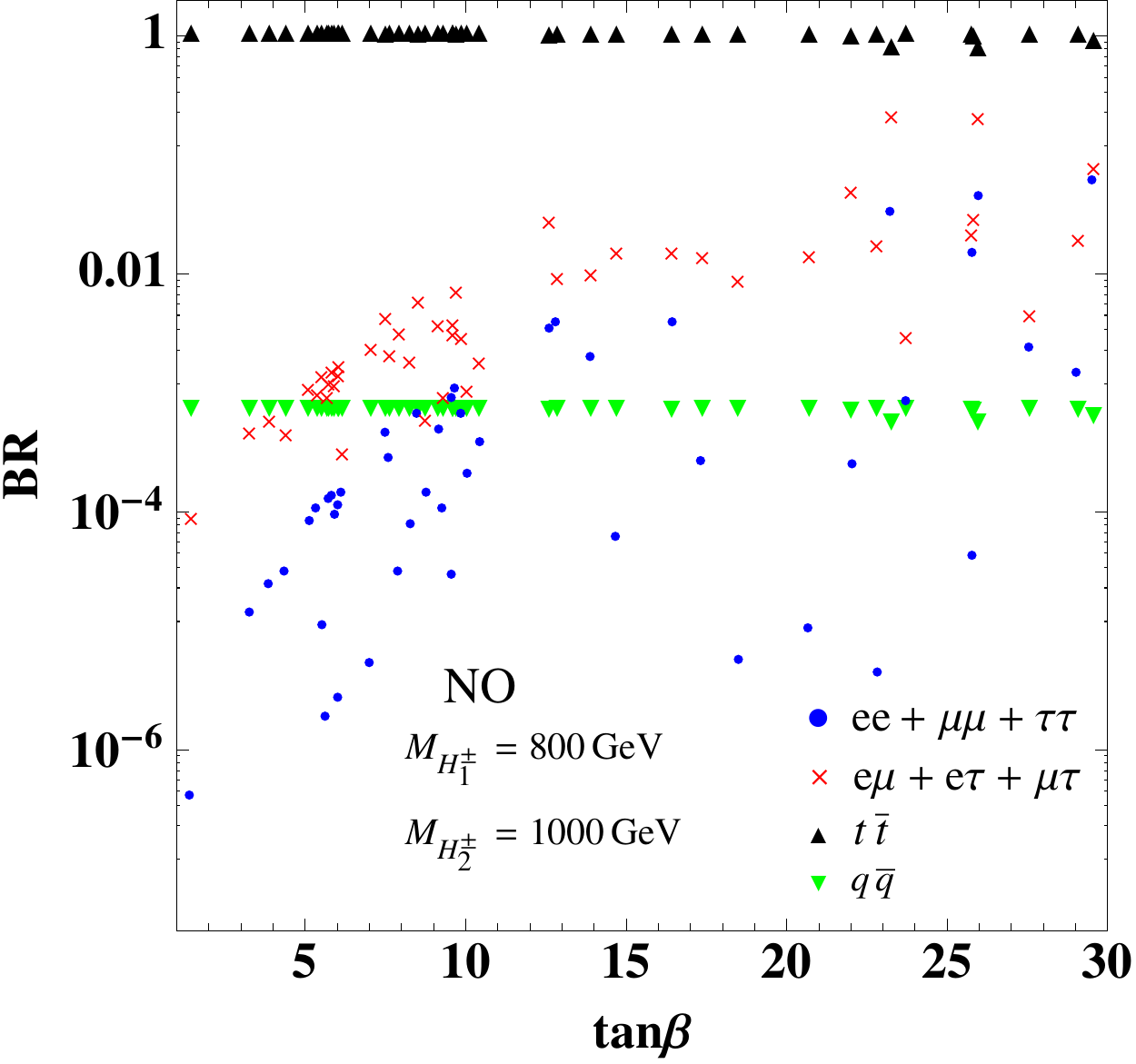} \qquad
\includegraphics[width=60mm]{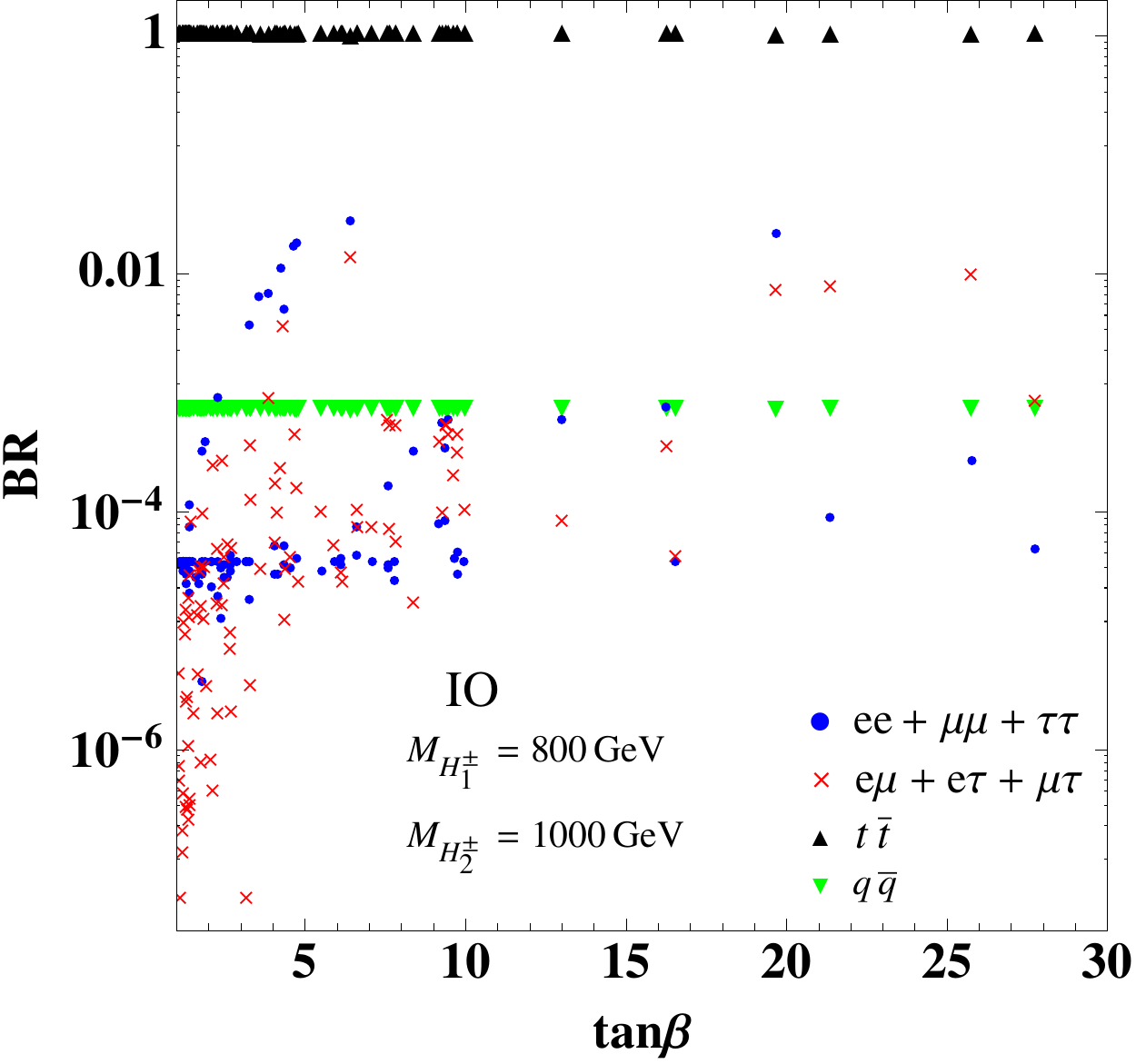}
\caption{BRs of $H$ for the lepton flavor conserving modes ($ee+\mu\mu+\tau\tau$), the LFV modes ($e\mu+e\tau+\mu\tau$), the hadronic modes $(\sum_{q\neq t} q \bar q)$ modes and 
the $t\bar{t}$ mode as a function of $\tan \beta$ in the NO case (left) and the IO case (right). 
The upper and lower panels show the case with ($m_{H_1^\pm}$, $m_{H_2^\pm}$) = (300 GeV, 500 GeV) and (800 GeV, 1000 GeV), respectively.  }   
\label{fig:BRH}\end{center}\end{figure}

Now, let us discuss the decay of the additional neutral Higgs boson $H$. 
The decay BRs of $A$ are almost the same as those of $H$ in our parameter configuration.
We impose the constraint from the CLFV decays.  
In Fig.~\ref{fig:BRH}, we show the sum of BRs for lepton flavor conserving modes, LFV modes and the hadronic modes (only the $t\bar{t}$ mode is separately shown) as a function of $\tan \beta$. 
We here take two sets of the charged Higgs boson masses, i.e., ($m_{H_1^\pm}$, $m_{H_2^\pm}$) = (300 GeV, 500 GeV) and (800 GeV, 1000 GeV) displayed in the upper and lower panels, respectively.
For the smaller mass case, we see that the leptonic decay modes can be dominant, particularly for the larger $\tan\beta$ region in the both NO and IO cases, 
because the decay rates of the hadronic modes are suppressed by $\cot^2\beta$. 
Similar behavior can also be seen in the Type-X THDM~\cite{Aoki:2009ha}, but it does not induce the LFV decays of $H$. 
We also see that the BRs of the LFV modes of $H$ are typically larger than the flavor conserving modes in the NO case. 
On the other hand, when we take the larger mass case
the BR of the $H\to t \bar t $ mode becomes dominant for the wide region of the parameter space, but 
it is slightly reduced for the larger $\tan\beta$ region. 
Again, the BR for the LFV modes is typically larger than the flavor conserving one in the NO case. 

\begin{figure}[t!]\begin{center}
\includegraphics[width=60mm]{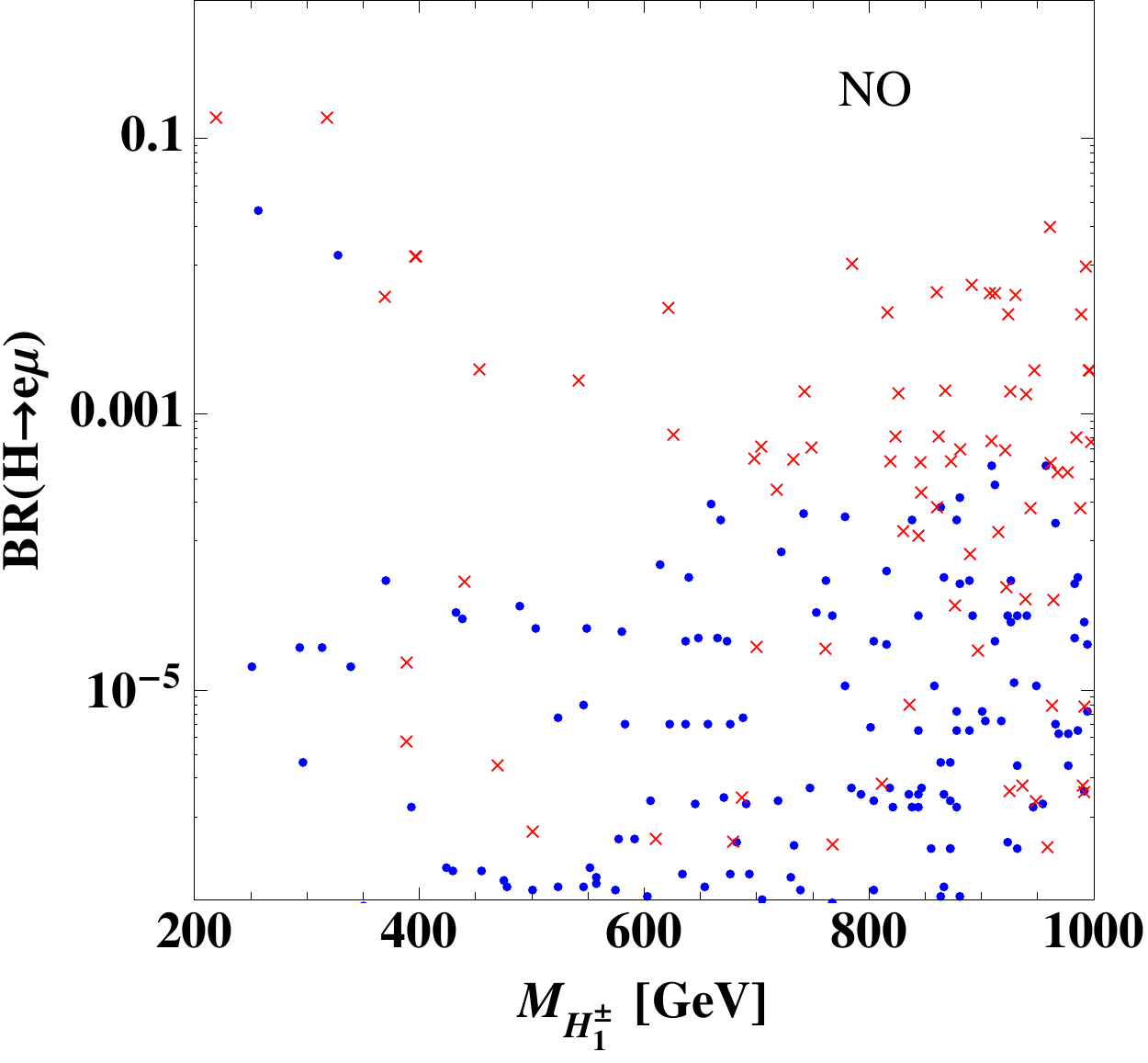} \qquad
\includegraphics[width=60mm]{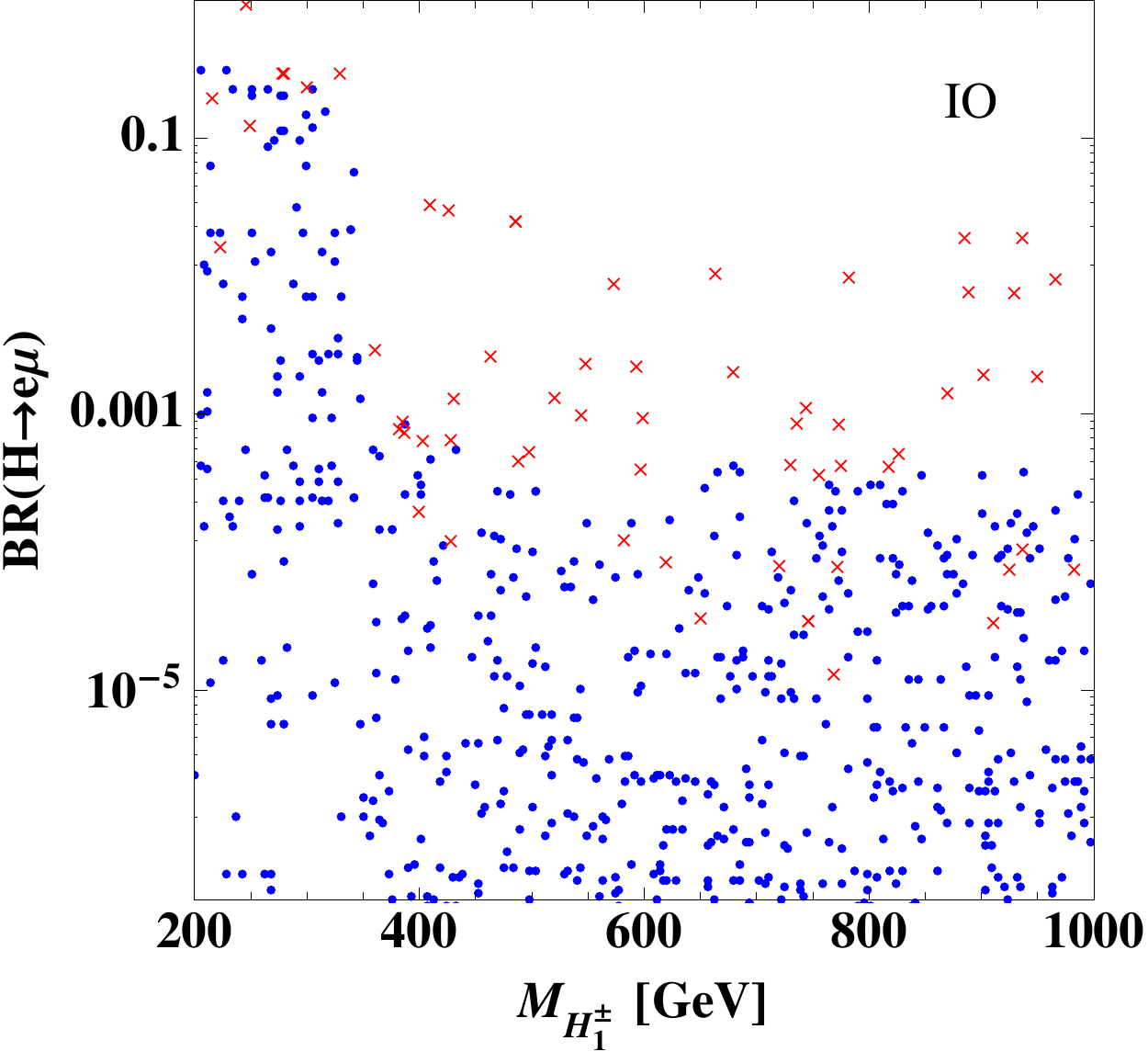} 
\includegraphics[width=60mm]{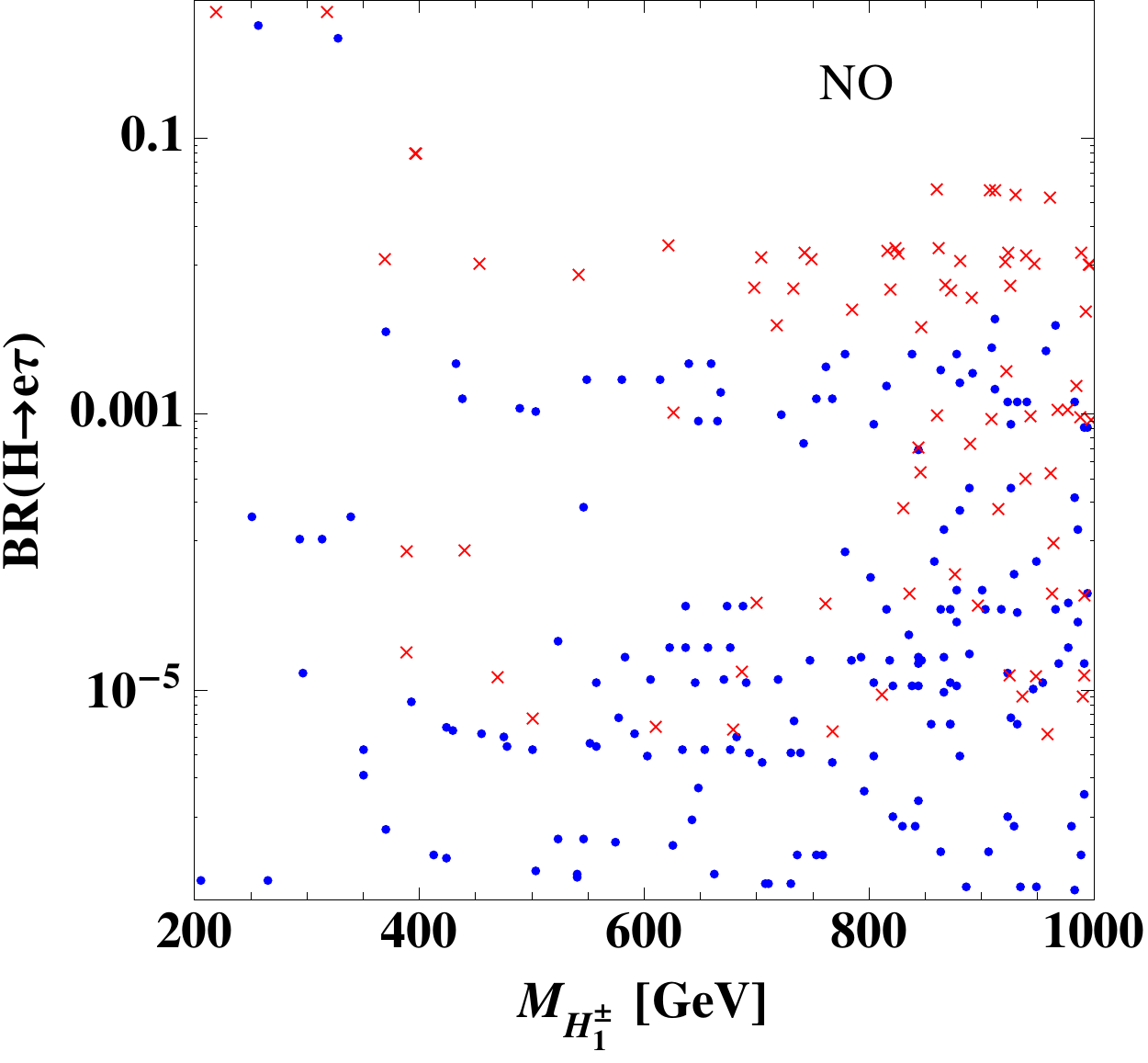} \qquad
\includegraphics[width=60mm]{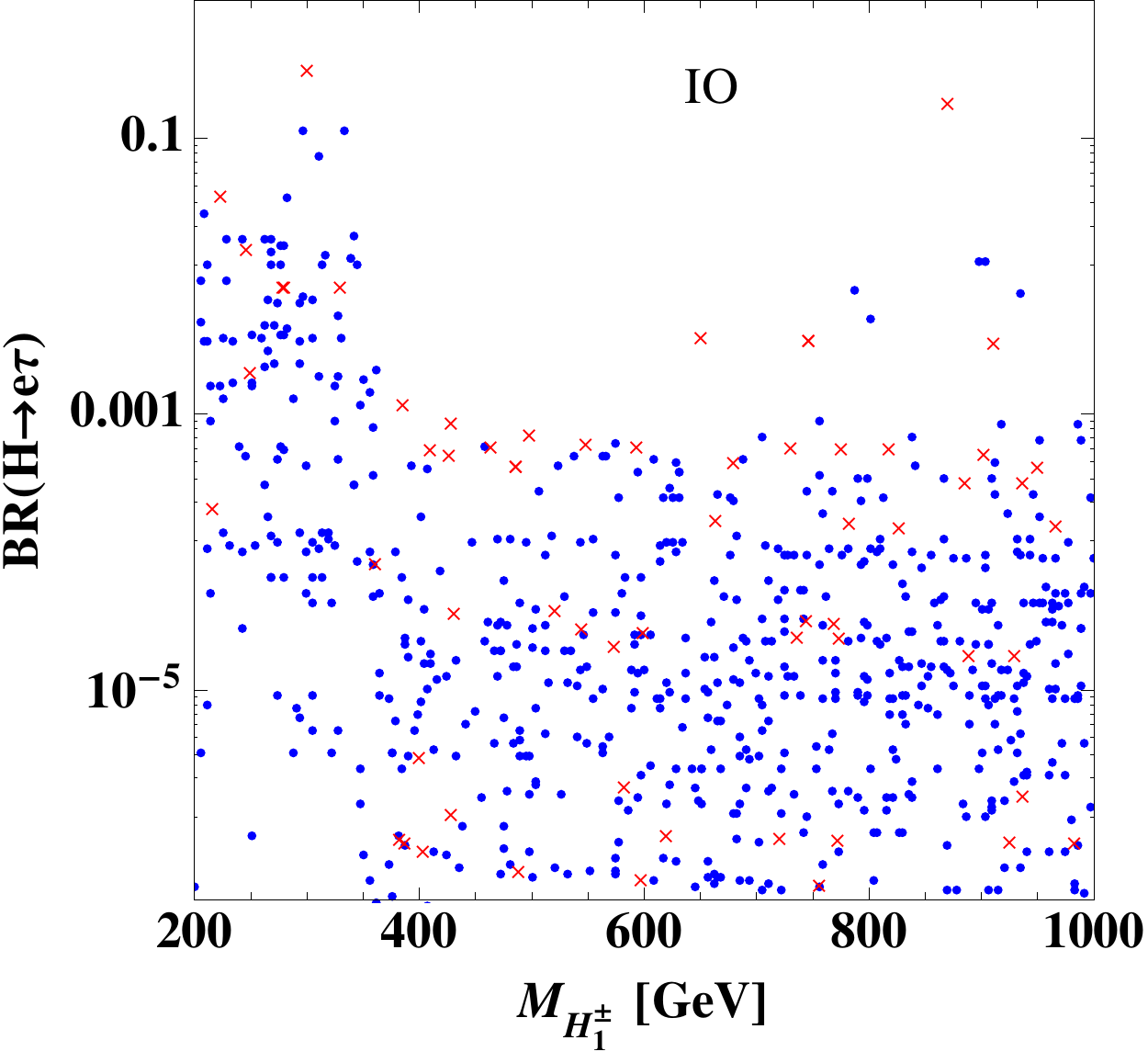} 
\includegraphics[width=60mm]{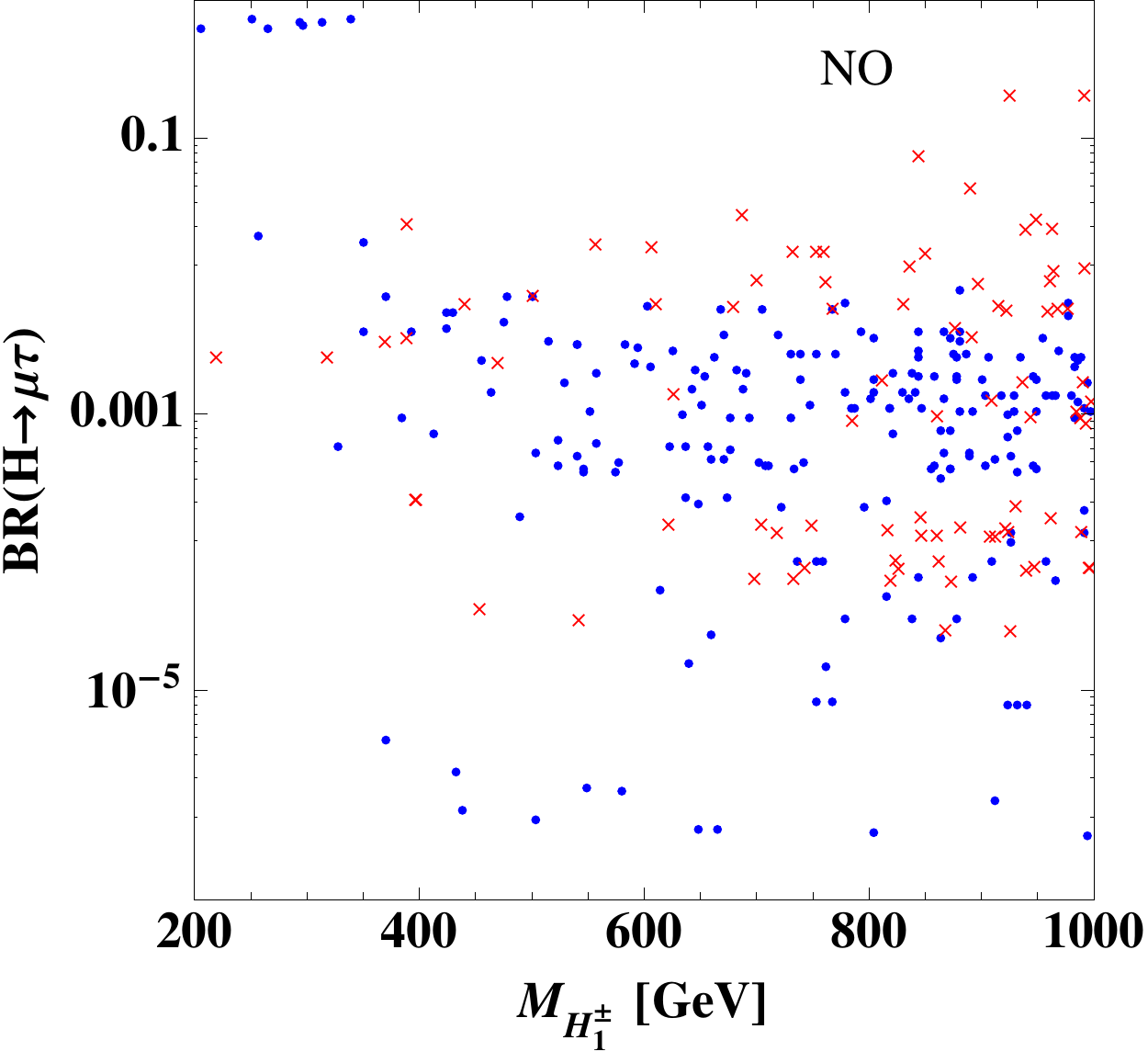} \qquad
\includegraphics[width=60mm]{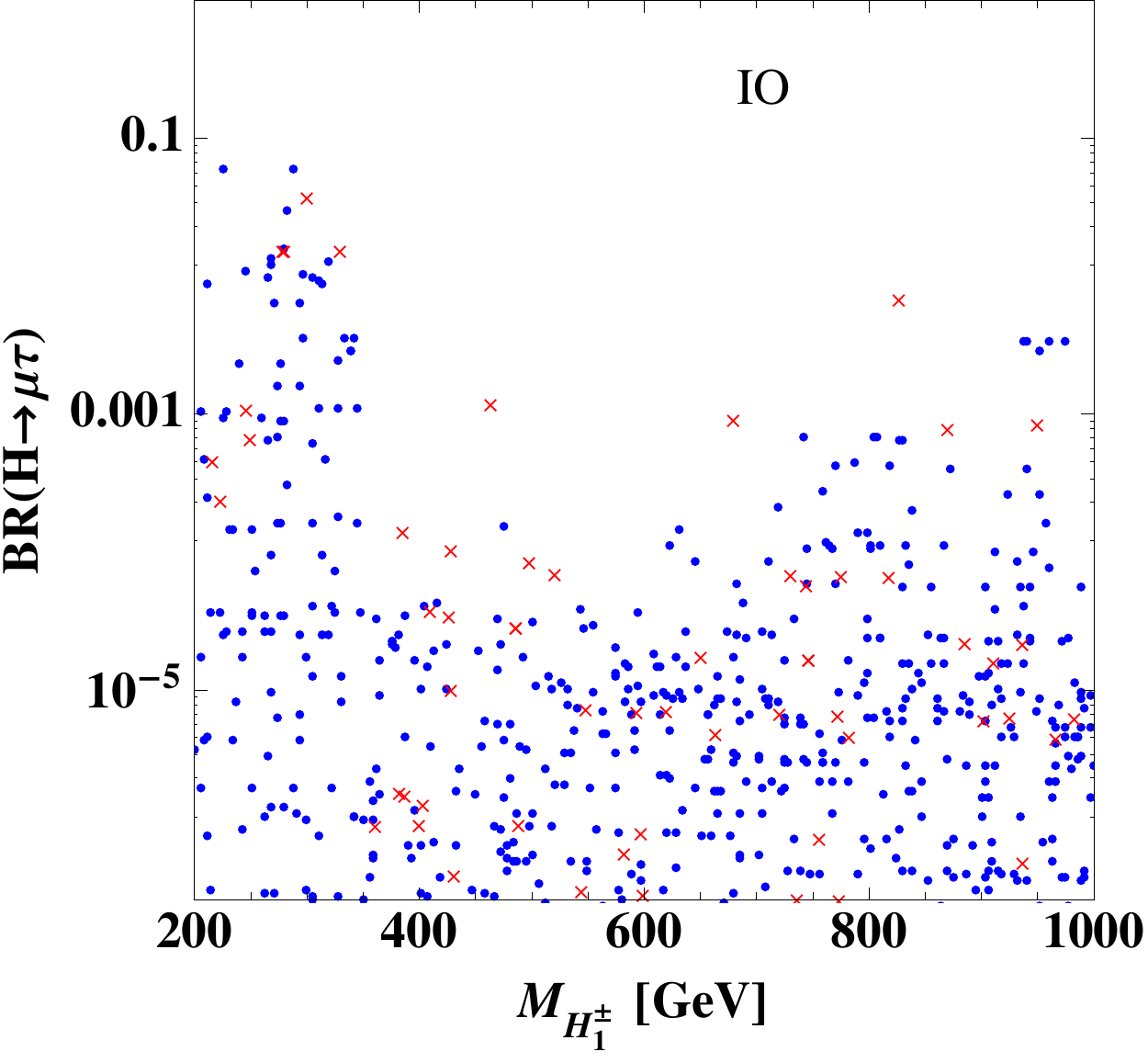}
\caption{BRs for LFV decay of $H$ as functions of $m_{H_1^\pm}^{}$ in the NO case (left) and the IO case (right). 
The blue dots (red crosses) show the case with $1< \tan\beta <10$ $(10 < \tan \beta < 30)$. 
}   
\label{fig:LFV3}\end{center}\end{figure}

Fig.~\ref{fig:LFV3} shows the BRs for the LFV decays of $H$ as a function of $m_{H_1^\pm}^{}$.
It is clearly seen that the BRs are suddenly suppressed at around $m_H = 2m_t$ due to the top pair threshold. 
As we already observed in Fig.~\ref{fig:BRH}, the case with a larger value of $\tan\beta$ has larger BRs for the LFV modes. 
For the mass region below $2m_t$, each BR can be tens percent, while for the larger mass region, the BRs 
of $H \to e\mu$ and $H \to e\tau$ can be maximally a few percent level. 
Only the BR of $H \to \mu\tau$ can be 10 percent level in the NO case even at $m_H^{} > 2m_t$, because this mode is less constrained by the CLFV decays.

\begin{figure}[t!]\begin{center}
\includegraphics[width=60mm]{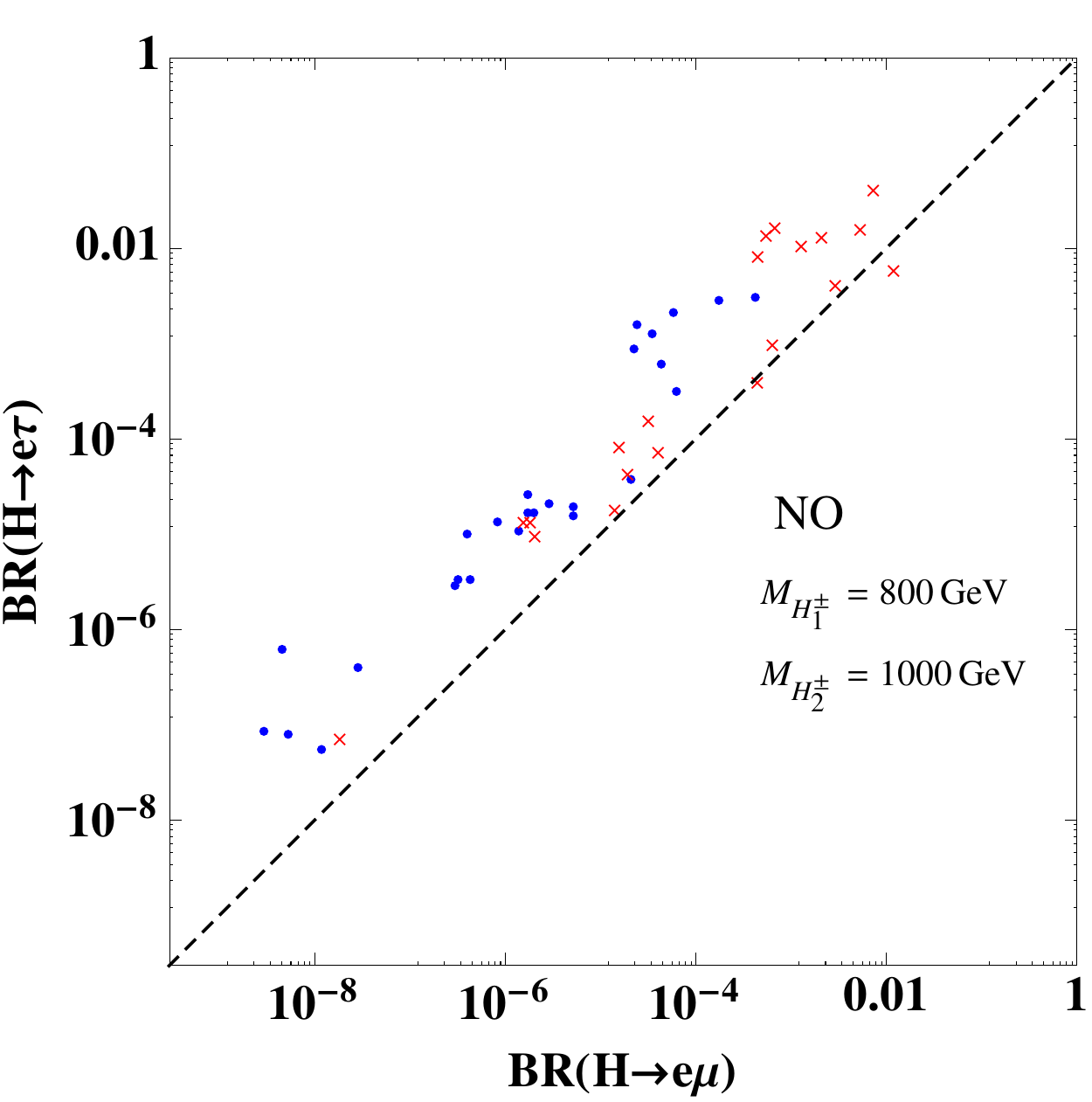} \qquad
\includegraphics[width=60mm]{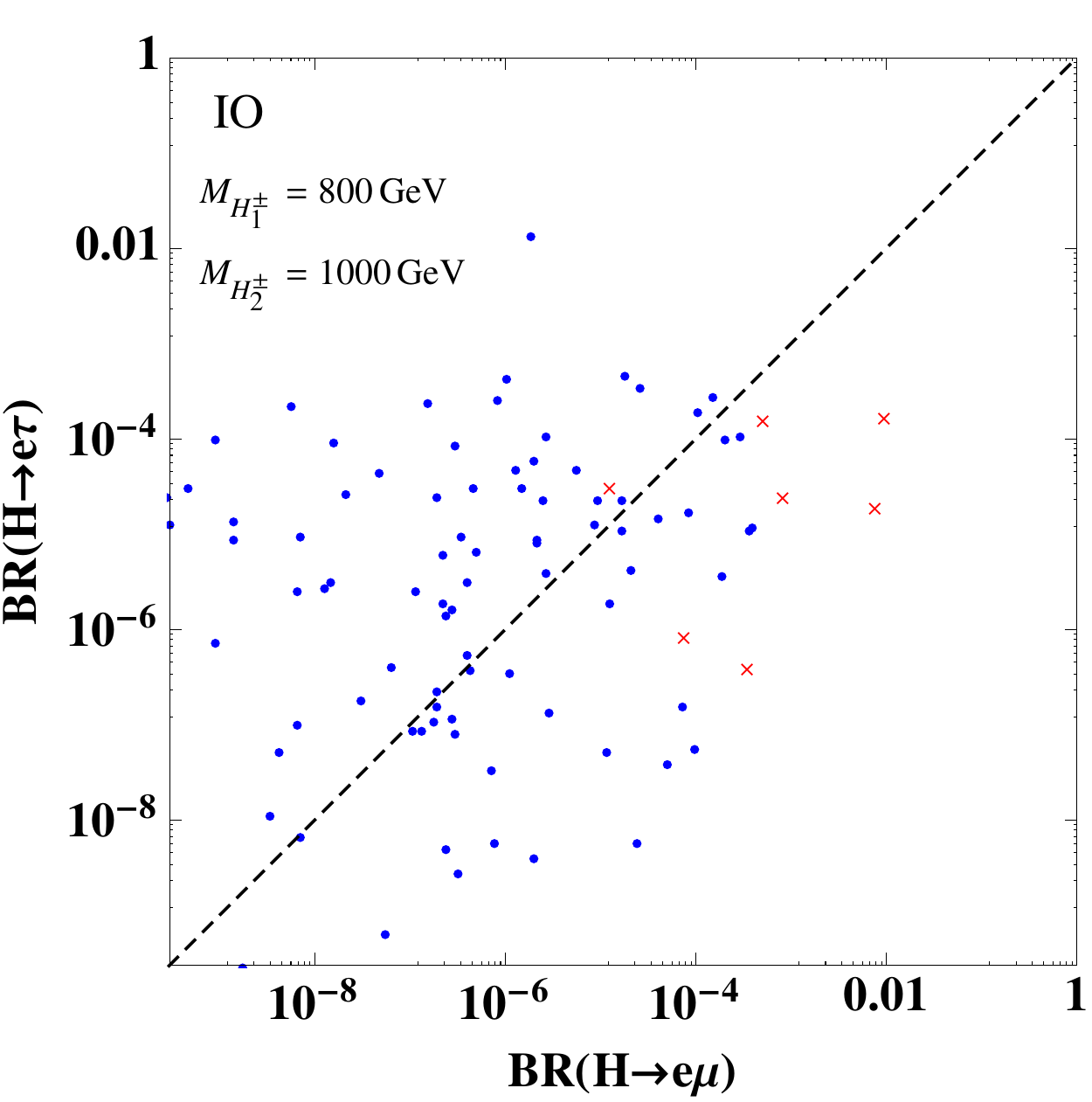}
\includegraphics[width=60mm]{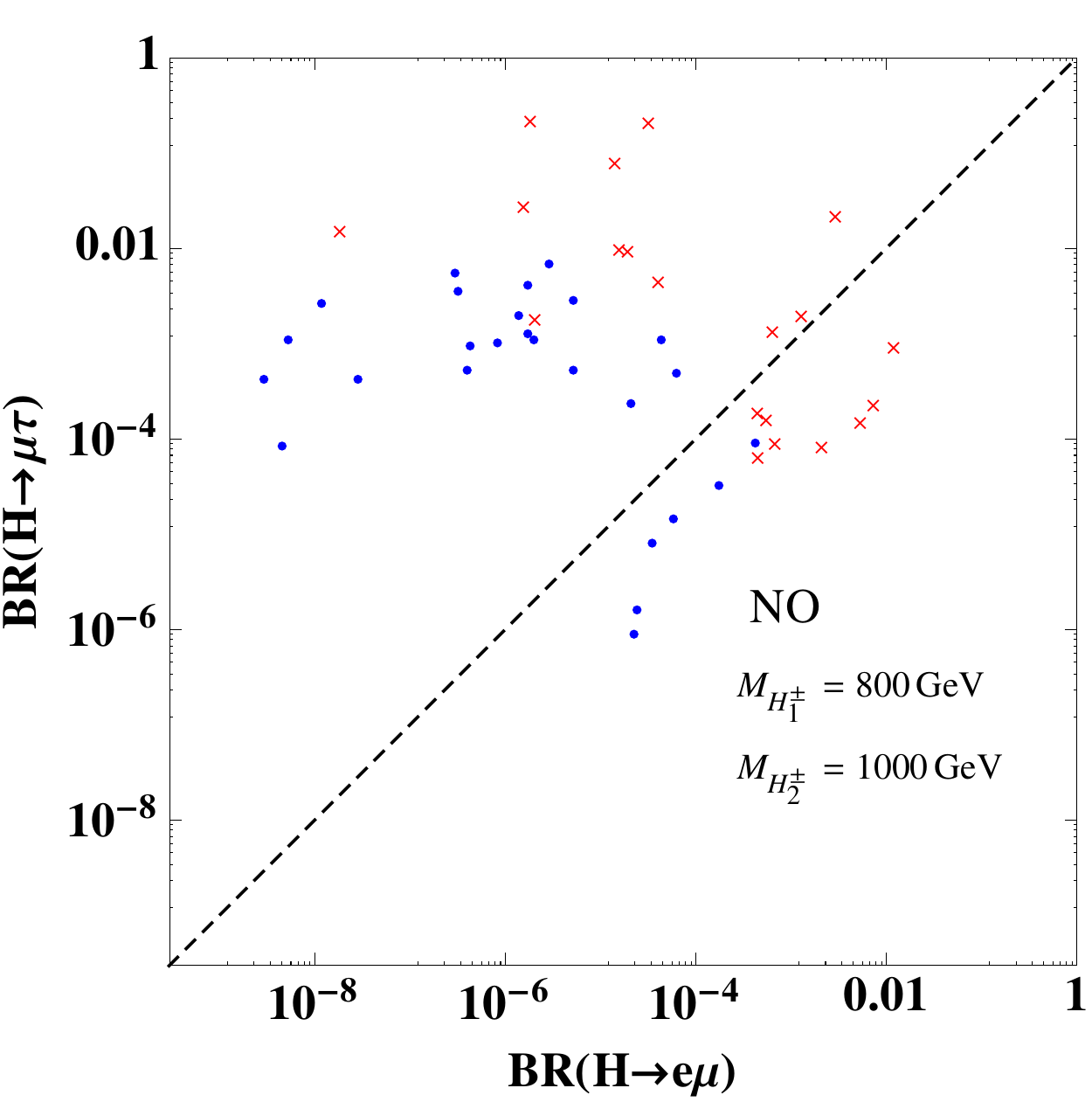} \qquad
\includegraphics[width=60mm]{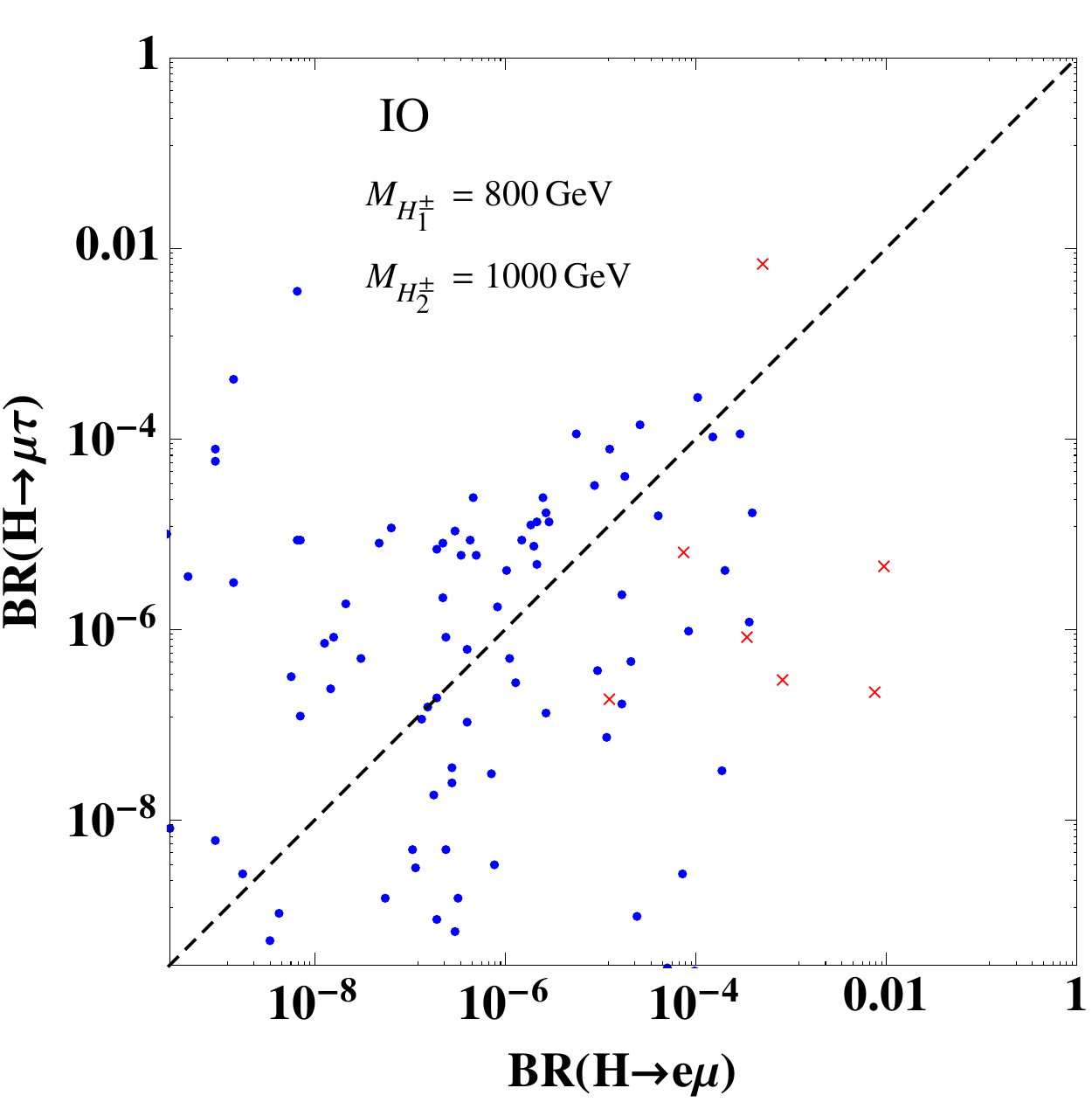}
\includegraphics[width=60mm]{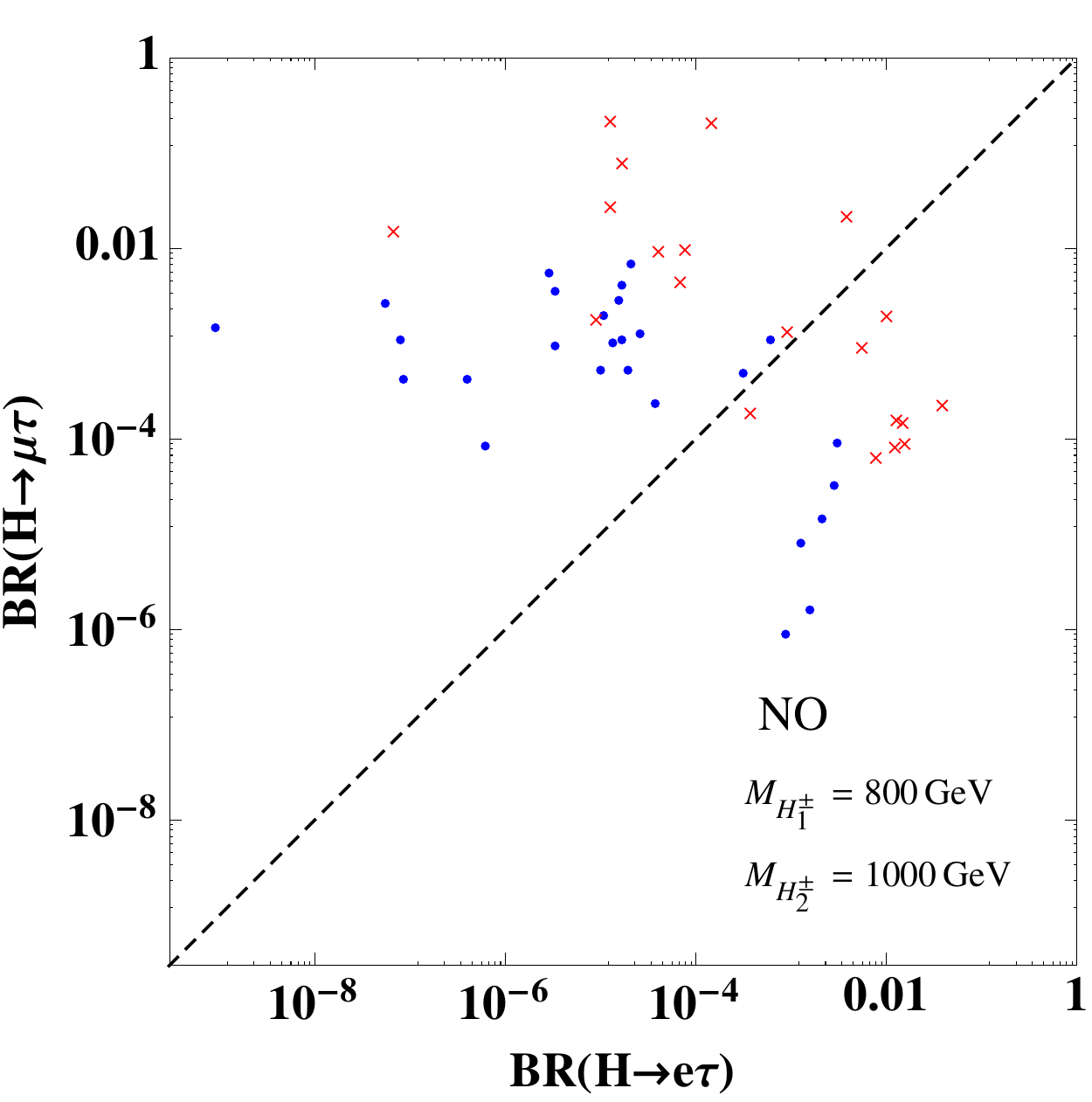} \qquad
\includegraphics[width=60mm]{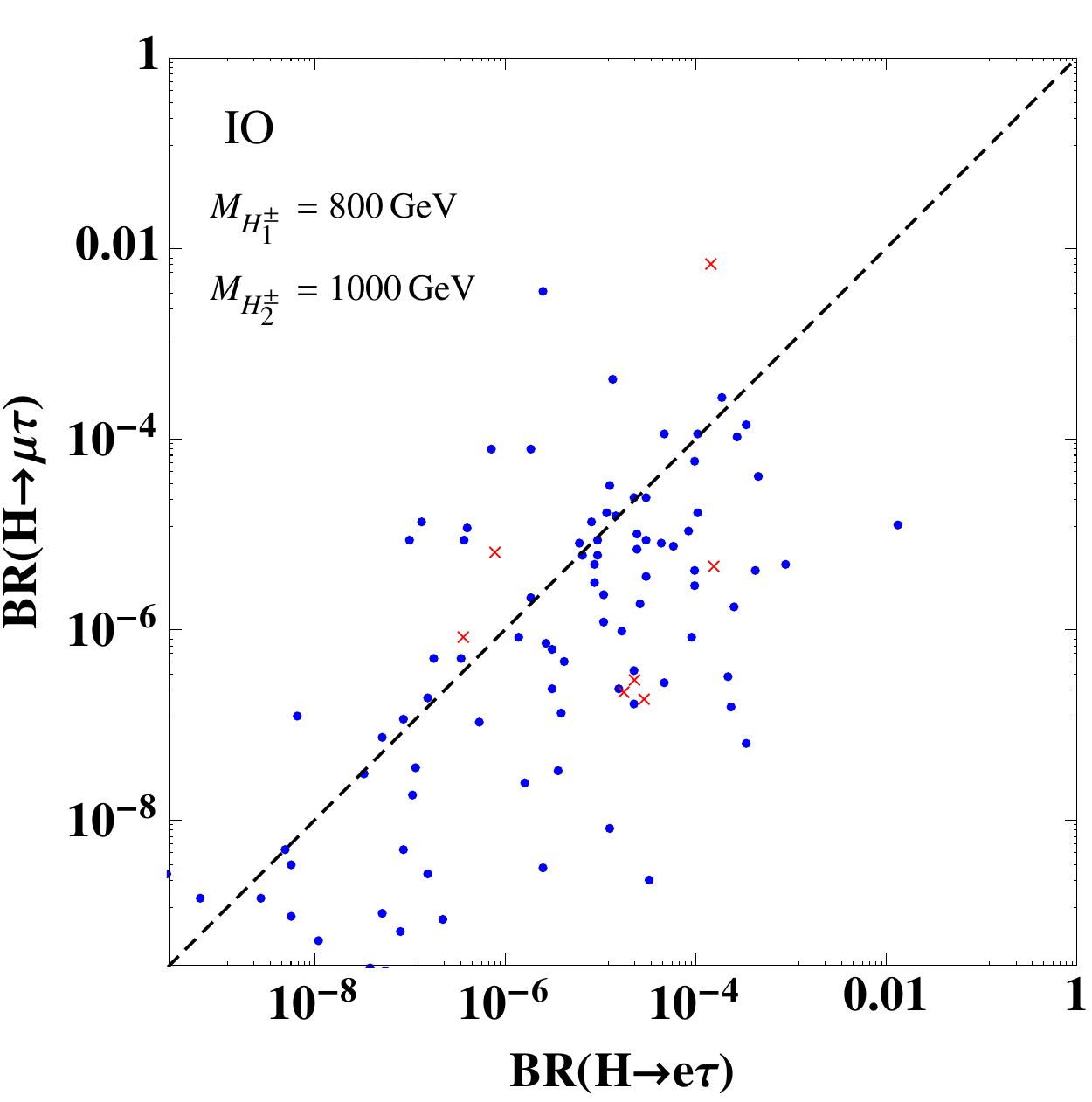}
\caption{Correlations among ${\rm BR}(H \to \ell_i \ell_j)$ in the NO case (left) and the IO case (right). 
The masses of charged Higgs bosons are fixed to be $m_{H_1^\pm}^{} = 800$ GeV and $m_{H_2^\pm}^{} = 1000$ GeV. 
The blue dots (red crosses) show the case with $1< \tan\beta <10$ $(10 < \tan \beta < 30)$.}   
\label{fig:LFV4}\end{center}\end{figure}

Finally in Fig.~\ref{fig:LFV4}, we show the correlations between two of three BRs by fixing the charged Higgs boson masses to be $m_{H_1^\pm}^{} = 800$ GeV and $m_{H_2^\pm}^{} = 1000$ GeV. 
These correlations do not change so much if we take the other values of these masses. 
In the NO case, we find the strong correlation between the $e\mu$ and $e\tau$ modes with ${\rm BR}(H \to e \tau) \gtrsim {\rm BR}(H \to e \mu)$, 
while the other correlations tends to be negative slightly.
On the other hand, in the IO case, BRs for $\mu \tau$ and $e \tau$ are positively correlated while the other correlations are not strong. 

These characteristic patterns of the additional neutral Higgs bosons can be tested at collider experiments. 
For example, at the LHC, we can use $pp \to \gamma^*/Z^* \to HA$ production whose production cross section is typically 
a few fb level with the masses of $H$ and $A$ to be 300 GeV at 13 TeV~\cite{Aoki:2009ha}. 
Therefore, we can expect order 100--1000 events of LFV decays of the additional neutral Higgs bosons with 3000 ab$^{-1}$ of the luminosity at the high-luminosity LHC. 
The $HH^\pm$ and $AH^\pm$ productions can also be useful together with the $pp \to HA$ mode. 
We note that the $gg \to H/A$ process  is not useful as the production mode, because the cross section is suppressed by $\cot^2\beta$, and we are 
mainly interested in the large $\tan\beta$ region, in which both $H$ and $A$ become lepton specific. 

Although we have not discussed the phenomenology of the singly-charged Higgs bosons $H_1^\pm$ and $H_2^\pm$, their collider signatures would also be important to prove our model. 
Detailed studies on searching for the singly-charged charged Higgs bosons have been done in Ref.~\cite{Cao:2017ffm} at the LHC and in Ref.~\cite{Cao:2018ywk} at future lepton colliders.

\section{Conclusion \label{sec:conc}}
 
We have discussed the simple extension of the Zee model, in which  
only the change is the replacement of a discrete $Z_2$ symmetry by a flavor dependent global $U(1)'$ symmetry. 
This simple modification makes the Zee model possible to explain the current neutrino oscillation data without introducing dangerous 
flavor changing neutral currents in the quark sector. 
We found a unique and successful charge assignment for the $U(1)'$ symmetry, i.e., Class I to explain the current neutrino oscillation data, 
where the right-handed lepton singlets and one of the Higgs doublets are charged under $U(1)'$. 
We then have shown the appearance of characteristic correlations of the lepton flavor violating decays of the charged leptons as well as the additional neutral Higgs bosons. 
In particular, we found that our model predicts the strong correlation, i.e., BR($H \to e\tau$) $\gtrsim$ BR($H \to e\mu$) in the normal ordering case for the neutrino mass hierarchy. 
By measuring such pattern of the Higgs boson decay, our model can be tested at collider experiments, and also distinguished from the usual two Higgs doublet models with a softly broken discrete $Z_2$ symmetry.

\begin{acknowledgments}

The work of KY is supported in part by the Grant-in-Aid for Early-Career Scientists, No.~19K14714. 

\end{acknowledgments}

\begin{appendix}

\section{Structure of lepton Yukawa matrices\label{sec:class}}

The charge assignments of the $U(1)'$ symmetry can be classified into three ways, named by Class I, Class II and Class III,  
where each of them is characterized by $(q_L^i=0,q_R^i \neq 0)$, $(q_L^i \neq 0,q_R^i = 0)$ and $(q_L^i \neq 0,q_R^i \neq 0)$, respectively.
We note that if we take the charge for $S^+$ denoted as $q_S^{}$ to be independent of $q_L^{i}$, then the matrix $\tilde{F}$ vanishes.
In this case, the singlet scalar $S^\pm$ does not carry the lepton number, and thus neutrinos are kept to be massless. 
Therefore, $q_S^{}$ should be determined by the choice of $q_L^{i}$. 
As we mentioned in Sec.~\ref{sec:model}, our $U(1)'$ symmetry can be replaced by a discrete $Z_3$ symmetry. 
We discuss how the same Lagrangian terms given in the $U(1)'$ symmetric model can be realized by imposing the $Z_3$ symmetry instead of $U(1)'$. 
In the following subsections, we discuss each class and the structure of the lepton Yukawa matrices in order. 

\subsection{Class I}

Class I is defined by the charge assignments 
\begin{align}
q_R = (0,0,-q),~~q_L = q_S = 0. \label{cl1}
\end{align}
We then obtain 
\begin{align}
\tilde{Y}_\ell^1 = \begin{pmatrix}
0 & 0 & \times \\
0 & 0 & \times \\
0 & 0 & \times
\end{pmatrix},\quad 
\tilde{Y}_\ell^2 = \begin{pmatrix}
\times & \times & 0 \\
\times & \times & 0 \\
\times & \times & 0
\end{pmatrix},\quad 
\tilde{F} = \begin{pmatrix}
0 & \times & \times \\
 & 0 & \times \\
 &  & 0
\end{pmatrix}, 
\end{align}
where $\times$ denotes a nonzero element, and the lower-left elements of $\tilde{F}$ are obtained by anti-symmetric nature of $\tilde{F}$. 
We note that we can construct the similar type of the matrix by assigning the $-q$ charge 
to the first or second element in $q_R^{}$ instead of the third element. 
In this case, the first or second column of $\tilde{Y}_\ell^1$ is filled by nonzero elements, and 
the other two columns of $\tilde{Y}_\ell^2$ are filled by nonzero elements. 
These choices, however, do not give physically different consequences from the first one defined in Eq.~(\ref{cl1}), 
so that we take the assignment given in Eq.~(\ref{cl1}) as the representative one for Class I. 

The matrix $\tilde{Y}_\ell$ defined in Eq.~(\ref{ye}) can be written in terms of the mass matrix for the charged leptons $M_\ell$ as 
\begin{align}
  \tilde{Y}_\ell = \frac{\sqrt{2}}{v}\left( \cot\beta M_\ell P_1 + \cot\beta M_\ell P_2 - \tan\beta M_\ell P_3 \right), 
\end{align}
where 
\begin{align}
P_1 = 
\begin{pmatrix}
1 & 0 & 0 \\
0 & 0 & 0 \\
0 & 0 & 0
\end{pmatrix},~~
P_2 = 
\begin{pmatrix}
0 & 0 & 0 \\
0 & 1 & 0 \\
0 & 0 & 0
\end{pmatrix},~~
P_3 = 
\begin{pmatrix}
0 & 0 & 0 \\
0 & 0 & 0 \\
0 & 0 & 1
\end{pmatrix}. \label{projection}
\end{align}

\subsection{Class II \label{sec:class2}}

Class II is defined by the charge assignments 
\begin{align}
q_L = (0,0,q),~~q_R = 0. \label{cl2}
\end{align}
Regardless of $q_S$, we obtain 
\begin{align}
\tilde{Y}_\ell^1 = \begin{pmatrix}
0 & 0 & 0 \\
0 & 0 & 0 \\
\times & \times & \times
\end{pmatrix},\quad 
\tilde{Y}_\ell^2 = \begin{pmatrix}
\times & \times & \times \\
\times & \times & \times \\
0 & 0 & 0
\end{pmatrix}.  
\end{align}
Because of the nonzero charge of $q_L^{}$, one or two independent elements of $\tilde{F}$ become zero depending on the choice of $q_S$. 
We have the following two choices: 
\begin{align}
\tilde{F} = \begin{pmatrix}
0 & \times & 0 \\
  & 0 & 0 \\
  &   & 0
\end{pmatrix}\quad \text{for }q_S = 0, \quad 
\tilde{F} = \begin{pmatrix}
0 & 0 & \times \\
  & 0 & \times \\
  &   & 0
\end{pmatrix}\quad \text{for }q_S = -q.   
\end{align}
The matrix $\tilde{Y}_\ell$  can be written by 
\begin{align}
  \tilde{Y}_\ell = \frac{\sqrt{2}}{v}\left( \cot\beta P_1M_\ell + \cot\beta P_2M_\ell - \tan\beta P_3M_\ell \right). 
\end{align}

\subsection{Class III}

Class III is defined by the charge assignments 
\begin{align}
q_L^i = (0,0,-q),~~q_R^i = (0,0,-2q). \label{cl3}
\end{align}
Regardless of $q_S$, we obtain 
\begin{align}
\tilde{Y}_\ell^1 = \begin{pmatrix}
0 & 0 & 0 \\
0 & 0 & 0 \\
0 & 0 & \times 
\end{pmatrix},\quad 
\tilde{Y}_\ell^2 = \begin{pmatrix}
\times  & \times  & 0 \\
\times  & \times  & 0 \\
0 & 0 & 0
\end{pmatrix}.  
\end{align}
The structure of $\tilde{F}$ depends on the charge $q_S$ as
\begin{align}
\tilde{F} = \begin{pmatrix}
0 & \times  & 0 \\
  & 0 & 0 \\
  &   & 0
\end{pmatrix}\quad \text{for }q_S = 0, \quad 
\tilde{F} = \begin{pmatrix}
0 & 0 & \times  \\
  & 0 & \times  \\
  &   & 0
\end{pmatrix}\quad \text{for }q_S = q.   
\end{align}
The matrix $\tilde{Y}_\ell$  can be written by 
\begin{align}
  \tilde{Y}_\ell = \frac{\sqrt{2}}{v}\left( \cot\beta \sum_{i,j}^{1,2} P_iM_\ell P_j - \tan\beta P_3M_\ell P_3\right). 
\end{align}

\subsection{Equivalence to the model with a $Z_3$ symmetry }

\begin{table}[!h]
\begin{center}
\begin{tabular}{|l||ccc|}\hline
          & $(L_L^1,L_L^2,L_L^3)$ & $(e_R^{},\mu_R^{},\tau_R^{})$ & $S^+$ \\\hline\hline
Class I   & (1,1,1) & (1,1,$\omega^2$) & 1   \\\hline
Class II  & (1,1,$\omega$) & (1,1,1) & 1 or $\omega$ \\\hline
Class III & (1,1,$\omega^2$) & (1,1,$\omega$) & 1 or $\omega^2$ \\\hline
\end{tabular}
\caption{$Z_3$ charge assignments for the leptons and the charged singlet $S^+$ to realize Class I, II and III. }
\label{particle2}
\end{center}
\end{table}

Let us here consider the model with a discrete $Z_3$ symmetry instead of the $U(1)'$ symmetry. 
We fix the $Z_3$ transformation of the two Higgs doublets as   
\begin{align}
\Phi_1 \to \omega  \Phi_1, \quad 
\Phi_2 \to  \Phi_2, 
\end{align}
where $\omega \equiv e^{2\pi i/3}$. All the quark fields are neutral under $Z_3$. 
The transformation property for the leptons and $S^\pm$ are shown in Table~\ref{particle2} depending on the classes which are defined in the above subsections. 
We note that the $Z_3$ symmetry is softly-broken by the $\Phi_1^\dagger \Phi_2$ and/or $\Phi_1^T(i\tau_2)\Phi_2S^-$ terms in the potential. 

\section{Mass formulae for the scalar bosons \label{sec:mass}}

The mass eigenvalues of the Higgs bosons are calculated as 
\begin{align}
m_A^2&=\frac{m_3^2}{s_\beta c_\beta},\\
m_H^2&=c^2_{\beta-\alpha}M_{11}^2 + s^2_{\beta-\alpha}M_{22}^2 - 2s_{\beta-\alpha}c_{\beta-\alpha}M_{12}^2,    \\
m_h^2&= s^2_{\beta-\alpha}(M_\text{even}^2)_{11} + c^2_{\beta-\alpha}(M_\text{even}^2)_{22} + 2s_{\beta-\alpha}c_{\beta-\alpha} (M_\text{even}^2)_{12} , \\
m_{H_1^\pm}^2 & = c^2_{\chi} (M_\pm^2)_{11} + s^2_{\chi} (M_\pm^2)_{22} + 2s_{\chi}c_{\chi} (M_\pm^2)_{12},    \\
m_{H_2^\pm}^2 & = s^2_{\chi} (M_\pm^2)_{11} + c^2_{\chi} (M_\pm^2)_{22} - 2s_{\chi}c_{\chi} (M_\pm^2)_{12} ,
\end{align}
where $(M_\text{even}^2)_{ij}$ and $(M_\pm^2)_{ij}$ ($i,j=1,2$) are the mass matrices for the CP-even and singly-charged Higgs bosons in the basis of $(h_1',h_2')$ and $(H^\pm, S^\pm)$, respectively. 
Each element is given as 
\begin{align}
(M_\text{even}^2)_{11}   & =v^2 \left(\lambda_1c^4_\beta+\lambda_2 s^4_\beta + \frac{\lambda_3+\lambda_4}{2} s^2_{2\beta} \right),\\
(M_\text{even}^2)_{22} & = m_A^2 + \frac{v^2}{8}[\lambda_1+\lambda_2-2(\lambda_3+\lambda_4)](1 - c_{4\beta}), \\
(M_\text{even}^2)_{12} & = \frac{v^2}{2}s_{2\beta} [-\lambda_1 c^2_\beta + \lambda_2 s^2_\beta + (\lambda_3+\lambda_4)c_{2\beta}],\\
(M_\pm^2)_{11} & = m_A^2-\frac{v^2}{2}\lambda_4, \\
(M_\pm^2)_{22} & = m_S^2+\frac{v^2}{2}(\sigma_1 c^2_\beta + \sigma_2 s^2_\beta), \\
(M_\pm^2)_{12} & = -v\frac{\mu}{\sqrt{2}},
\end{align}
The mixing angles are expressed in terms of these matrix elements: 
\begin{align}
\tan2(\alpha - \beta) &=\frac{2(M_{\text{even}}^2)_{12}}{(M_{\text{even}}^2)_{11} - (M_{\text{even}}^2)_{22}}, \\
\tan2\chi             &=\frac{2(M_\pm^2)_{12}}{(M_\pm^2)_{11} - (M_\pm^2)_{22}}. 
\end{align}

\section{Amplitudes for $\ell_i \to \ell_j \gamma$ \label{sec:LFV_amp}}

We present the analytic formulae for the amplitudes of the $\ell_i \to \ell_j \gamma$ processes denoted by $(a_{L,R})_{ij}$ in Eq.~(\ref{br_clfv}). 
The contributions from the charged Higgs boson loops are given by  
\begin{align}
(a_{R}^{H^\pm_1})_{ij} &= \frac{1}{16 \pi^2} \sum_{k=1}^3 \left[ (Y_\ell)_{kj}^{*} (Y_\ell)_{ki} c_\chi^2 F_1(m_{\ell_i}, m_{\ell_j}, m_{H_1^\pm})- F_{kj}^* F_{ki} s_\chi^2 F_2(m_{\ell_i}, m_{\ell_j}, m_{H_2^\pm})   \right], \\
(a_{L}^{H^\pm_1})_{ij} &= \frac{1}{16 \pi^2} \sum_{k=1}^3 \left[ (Y_\ell)_{kj}^{*} (Y_\ell)_{ki} c_\chi^2 F_2(m_{\ell_i}, m_{\ell_j}, m_{H_1^\pm})- F_{kj}^* F_{ki} s_\chi^2 F_1(m_{\ell_i}, m_{\ell_j}, m_{H_2^\pm})   \right], \\
(a_{R}^{H^\pm_2})_{ij} &= (a_{R}^{H^\pm_1})_{ij}\Big|_{c_\chi^2 \leftrightarrow s_\chi^2}, \quad (a_{L}^{H^\pm_2})_{ij} = (a_{L}^{H^\pm_1})_{ij}\Big|_{c_\chi^2 \leftrightarrow s_\chi^2}. 
\end{align}
The loop functions are written as 
\begin{equation}
F_{1[2]} (m_1, m_2, m_3) = \int [dX] \frac{x z m_2 [x y m_1]}{(x^2-x) m_1^2 + x z (m_1^2 - m_2^2)  + (y+z) m_3^2},
\end{equation}
where $\int [dX] \equiv \int_0^1 dx dy dz \delta(1-x-y-z) $. 
Similarly, the contributions from the neutral scalar boson ($\varphi = h, H, A$) loops are given by
\begin{align}
& (a_{R}^\varphi)_{ij} = \frac{1}{8 \pi^2} \sum_{k=1}^3 \int [dX] \frac{xy m_{\ell_i} f^{jk}_\varphi f^{ki}_\varphi + x z m_{\ell_j} g^{jk}_\varphi g^{ki}_\varphi + (1-x) m_{\ell_k} f^{jk}_\varphi g^{ki}_\varphi }{-x(1-x) m^2_{\ell_i} - x z (m^2_{\ell_j} - m^2_{\ell_i}) + (z+ y) m^2_{\ell_k} + x m_\varphi^2}, \\
& (a_{L}^\varphi)_{ij} = \frac{1}{8 \pi^2} \sum_{k=1}^3 \int [dX] \frac{xz m_{\ell_i} f^{jk}_\varphi f^{ki}_\varphi + x y m_{\ell_j} g^{jk}_\varphi g^{ki}_\varphi + (1-x) m_{\ell_k} g^{jk}_\varphi f^{ki}_\varphi }{-x(1-x) m^2_{\ell_i} - x z (m^2_{\ell_j} - m^2_{\ell_i}) + (z+ y) m^2_{\ell_k} + x m_\varphi^2}, 
\end{align}
where the couplings for the scalars $\varphi$ are defined as 
\begin{align}
& f^{ij}_h = \frac{\sqrt{2} m_{\ell_i}}{v} s_{\beta-\alpha} \delta_{ij} +  \frac{1}{\sqrt{2}} (Y^0_\ell)^*_{ji} c_{\beta-\alpha}, 
\quad g^{ij}_h = \frac{\sqrt{2} m_{\ell_i}}{v} s_{\beta-\alpha} \delta_{ij} + \frac{1}{\sqrt{2}} (Y^0_\ell)_{ij} c_{\beta-\alpha} \nonumber \\
& f^{ij}_H = \frac{\sqrt{2} m_{\ell_i}}{v} c_{\beta-\alpha} \delta_{ij} - \frac{1}{\sqrt{2}} (Y^{0}_\ell)^*_{ji} s_{\beta-\alpha}, 
\quad  g^{ij}_H = \frac{\sqrt{2} m_{\ell_i}}{v} c_{\beta-\alpha} \delta_{ij} - \frac{1}{\sqrt{2}} (Y^0_\ell)_{ij} s_{\beta-\alpha}  \nonumber \\
& f^{ij}_A = - \frac{i}{\sqrt{2}} (Y^{0}_\ell)^*_{ji} c_{\beta-\alpha}, \quad g^{ij}_A = \frac{i}{\sqrt{2}} (Y^0_\ell)_{ij} c_{\beta-\alpha}.
\end{align}

\end{appendix}

\bibliography{references}

\end{document}